\documentclass[a4paper, 11pt]{article}
		
\usepackage{graphicx}
\usepackage{float}
\usepackage[utf8]{inputenc}
\usepackage{amsmath}
\usepackage{amsfonts}
\usepackage{color}
\usepackage{enumerate}
\usepackage{bm}
\usepackage{url}
\usepackage{subfigure}

\newcommand{\R}{\mathbb{R}}

\newcommand{\C}{\mathbb{C}}

\DeclareMathOperator{\diag}{diag}
\DeclareMathOperator{\rank}{rank}
\DeclareMathOperator{\spa}{span}
\DeclareMathOperator{\colsp}{colsp}

\newcommand{\bb}{\mathbf{b}}
\newcommand{\hb}{\mathbf{h}}
\newcommand{\kb}{\mathbf{k}}
\newcommand{\ub}{\mathbf{u}}
\newcommand{\vb}{\mathbf{v}}

\newcommand{\xb}{\mathbf{x}}
\newcommand{\yb}{\mathbf{y}}
\newcommand{\zb}{\mathbf{z}}
\newcommand{\alphab}{\bm{\alpha}}
\newcommand{\betab}{\bm{\beta}}
\newcommand{\etab}{\bm{\eta}}
\newcommand{\deltab}{\bm{\delta}}
\newcommand{\Hb}{\mathbf{H}}
\newcommand{\Lb}{\mathbf{L}}
\newcommand{\Ib}{\mathbf{I}}
\newcommand{\Mb}{\mathbf{M}}

\newtheorem{thm}{Theorem}

\newtheorem{prop}{Proposition}

\begin{document}

\title{\bf An Arrow-Hurwicz-Uzawa Type Flow as Least Squares Solver for Network Linear Equations}

\author{Yang Liu, Christian Lageman, Brian D. O. Anderson, and  Guodong Shi\thanks{Y. Liu, B. D. O. Anderson, and   G. Shi are with the Research School of Engineering, The Australian National University, ACT 0200,
Canberra, Australia. C. Lageman is with Institute of Mathematics, University of W\"{u}rzburg,  W\"{u}rzburg 97070,
Germany.  Email: yang.liu@anu.edu.au, christian.lageman@mathematik.uni-wuerzburg.de,  brian.anderson@anu.edu.au, guodong.shi@anu.edu.au}
}
\date{}
\maketitle
\begin{abstract}
We study the approach to obtaining least squares solutions to systems of linear algebraic equations over networks by using distributed algorithms. Each node has access to one of the linear equations and holds a dynamic state. The aim for the node states is to reach a consensus as a least squares solution of the linear equations by exchanging their states with neighbors over an underlying interaction graph. A continuous-time distributed least squares solver over networks is developed in the form of the famous Arrow-Hurwicz-Uzawa flow. A necessary and sufficient condition is established on the graph Laplacian for the continuous-time distributed algorithm to give the least squares solution in the limit, with an exponentially fast convergence rate. The feasibility of different fundamental graphs is discussed including path graph, star graph, etc. Moreover, a discrete-time distributed algorithm is developed by Euler's method, converging exponentially to the least squares solution at the node states with suitable step size and graph conditions. The exponential convergence rate for both the continuous-time and discrete-time algorithms under the established conditions is confirmed by  numerical examples. Finally, we investigate the performance of the proposed flow under switching networks, and surprisingly,   switching networks at high switching frequencies can lead to approximate least square solvers even if all graphs in the switching signal fail to do so in the absence of structure switching.
\end{abstract}

Keywords: Distributed Algorithms; Dynamical Systems; Linear Equations; Least Squares; Switching Networks.	

\section{Introduction}
Linear algebraic equations are foundational for various  computation tasks, and   systems of linear algebraic equations also arise from various practical engineering problems \cite{garland2008parallel,keckler2011gpus,partl2011enabling,preparata1981cube,elbirt2005instruction,ayari2016multi,de2007distributed}. In recent years much interest has developed in finding out how to solve linear equations using multiple processing units or over a network. Major efforts have been made in the development of parallel or  distributed algorithms as linear-equation solvers. On the one hand one aims at faster algorithms in view  of the intuition that multiple processing units working in parallel under smart arrangements might provide significant improvements in computation efficiency. On the other hand various distributed systems induce structure constraints, e.g., one node holds one equation and it cannot or does not want to share the exact equation with other nodes, a constraint which excludes the feasibility of classical centralized algorithms.

Parallel algorithms for linear equations  have been developed  in the spirit of high-performance computing, e.g., the Jacobi method \cite{margaris2014parallel,yang2014acceleration}, the successive over-relaxations method \cite{young1954iterative}, the Kaczmarz method \cite{kaczmarz1937angenaherte} and the randomized iterative method \cite{gower2015randomized}. In these algorithms, the state of each node can give an entry of the solution to a linear equation after a suitably long running  time, via successive information exchange with other nodes and parallel independent computing. There are two restrictions in these parallel algorithms. The first restriction is that often each node is implicitly required to have access to all the other ones, i.e., the network graph is naturally complete \cite{margaris2014parallel,yang2014acceleration,kaczmarz1937angenaherte,gower2015randomized}. Second, linear equations are restricted in many parallel algorithms. It is somewhat unsatisfactory that, for example in the Jacobi method,  a sufficient condition for convergence is that the linear equations must be strictly or irreducibly diagonally dominant.

Meanwhile, discrete and continuous-time algorithms for linear equations known to have a unique solution are also established from the point of view of  distributed control and optimization. A variety of distributed algorithms are presented, among which discrete-time algorithms are given by \cite{mou2013fixed,mou2015distributed,liu2013asynchronous,lu2009distributed1,lu2009distributed2} and continuous-time algorithms are presented in \cite{anderson2015decentralized,shi:15}. In these network distributed algorithms, compared with the development in parallel computing, each node state asymptotically converges to the solution to the linear equation. Such developments on network linear equations are closely related to, and sometimes even special cases of,  the study of distributed optimization methods on nonlinear models \cite{tsitsiklis1984distributed,nedic2009distributed1,nedic2010constrained1,gharesifard2014distributed,jakovetic2014fast}, due to the natural connection between solving equations and optimizing objective functions.

Most of the existing work for parallel and distributed algorithms assumes that the linear equations have exact solutions
\cite{margaris2014parallel,yang2014acceleration,young1954iterative,kaczmarz1937angenaherte,gower2015randomized,mou2013fixed,mou2015distributed,liu2013asynchronous,lu2009distributed1,lu2009distributed2,anderson2015decentralized}, or can only produce least square solutions in the approximate sense or for limited graph structures \cite{shi:15}, and very few results have been obtained on exact distributed least squares solvers for network linear equations.  In this paper, distributed continuous and discrete-time algorithms that can compute the least squares solution to a linear equation over a network are presented. The particular contributions of our work are summarized as follows.
\begin{itemize}
\item  By recognizing a least-squares problem for a linear equation as a constrained optimization problem over a network,  a continuous-time flow  is presented in the form of the classical Arrow-Hurwicz-Uzawa flow \cite{arrow:58}, for which  we establish necessary and sufficient conditions regarding which graph structures can drive  the network flow to agree on the least squares solution.

\item By Euler's method, a discrete-time algorithm is presented and the properties of its convergence are also specified and proved.

 \item Comprehensive  discussions on the residual vector, an alternative state expansion method, and switching networks are provided. Surprisingly, it can be shown that under a sufficiently fast switching signal, the proposed flow for switching networks becomes an approximate least square solver under which node states are driven to at least a small neighborhood around the least square solution, even if all graphs in the switching signal fail to do so in the absence of structure switching.
\end{itemize}

The paper begins by the formulating the network linear equations in Section \ref{sec:2}, in addition to explaining how the Arrow-Hurwicz-Uzawa flow can be used to derive a continuous-time network flow. In Section \ref{sec:3}, a necessary and sufficient condition for the continuous-time flow to converge to the least squares solution is established, along with its proof and a couple of numerical examples. In Section \ref{sec:4}, a discrete-time algorithm is obtained by Euler's method and the necessary and sufficient conditions for its convergence conditions are established, and a few numerical examples are also given. In Section \ref{sec:5}, we further discuss  distributed computation of residual vector, alternative methods of obtaining the least squares solution, and a few numerical examples to study the convergence properties of the continuous-time algorithm over a switching network. Finally a few concluding remarks are given in Section \ref{sec:6}.

\section{Problem Definition} \label{sec:2}

\subsection{Linear Equation}
Consider the following linear algebraic equation with unknown $\yb\in \R^m$:
\begin{equation} \label{eq:linear_equation}
\zb = \Hb \yb
\end{equation}
and where $\zb\in\R^N$ and $\Hb\in\R^{N\times m}$ are known. Denote the column space of a matrix $\Mb$ by $\colsp\{\Mb\}$. If $\zb\in\colsp\{\Hb\}$, then the equation (\ref{eq:linear_equation}) always has (one or many) exact solutions. If $\zb\notin\colsp\{\Hb\}$, the least squares solution is defined by the solution of the following optimization problem:
\begin{equation}\label{eq:leastSquare_constaint}
\min_{\yb\in\R^m} \|\zb-\Hb\yb\|^2.
\end{equation}
It is well known that if $\rank(\Hb)=m$, then (\ref{eq:leastSquare_constaint}) yields a unique solution $\yb^\ast=(\Hb^\top \Hb)^{-1}\Hb^\top\zb$.

\subsection{Network}
Denote
\[
H =
\begin{bmatrix}
\hb_1^\top \\
\hb_2^\top \\
\vdots \\
\hb_N^\top
\end{bmatrix},
\quad
\zb =
\begin{bmatrix}
z_1 \\
z_2 \\
\vdots \\
z_N
\end{bmatrix}.
\]
We can rewrite (\ref{eq:linear_equation}) as
\[
\hb_i^\top \yb = z_i,\ i=1,\dots,N.
\]
Let $\mathcal{G}=(\mathcal{V},\mathcal{E})$ be a constant, undirected and connected graph with the set of nodes $\mathcal{V}=\{1,2,\dots,N\}$ and the set of edges $\mathcal{E}\subset\mathcal{V}\times\mathcal{V}$. Each node $i$ holds the equation $\hb_i^\top\yb=z_i$ and also holds a vector $\xb_i(t)\in\mathbb R^m$ that varies as a function of time $t$. Note that $\xb_i(t)$ will turn out to be part of the state of node $i$ at time $t$. Let $\mathcal{N}_i$ be the set of neighbor nodes that are connected to node $i$, i.e., $\mathcal{N}_i=\{j:(i,j)\in\mathcal{E}\}$. Define a diagonal matrix $\mathbf{D}=\diag(|\mathcal{N}_1|, |\mathcal{N}_2|, \dots, |\mathcal{N}_N|)$ and an incidence matrix $\mathbf{A}$ of the graph $\mathcal{G}$ by $[\mathbf{A}]_{ij}=1$ if $(i,j)\in\mathcal{E}$ and $[\mathbf{A}]_{ij}=0$ otherwise. Then $\mathbf{L}=\mathbf{D}-\mathbf{A}$ is the Laplacian of graph $\mathcal{G}$.

\subsection{Distributed Flows}
Consider a cost function $U(\cdot): \R^m\times\dots\times\R^m \to \R$
\begin{equation}\label{eq:U}
U(\xb_1,\dots,\xb_N) = \sum_{i=1}^{N}|\hb_i^\top\xb_i-z_i|^2.
\end{equation}
Let $\xb(t)=[\xb_1^\top(t) \ \dots \ \xb_N^\top(t)]^\top$ and introduce $\vb(t)=[\vb_1^\top(t) \ \dots \ \vb_N^\top(t)]^\top$ with $\vb_i(t)\in\R^m$ for $i=1,\dots,N$. The vector $\vb_i(t)$ is also held by node $i$, and the $2m$-dimensional vector $[\xb_i(t)^\top \ \vb_i(t)^\top]^\top$ represents the state of node $i$.

We consider the following flow:
\begin{equation}
\begin{aligned}
\dot{\xb} &=  - (\Lb\otimes \Ib_m) \vb - \nabla U(\xb) \\
\dot{\vb} &= (\Lb\otimes \Ib_m) \xb. \label{eq:flowModel1}
\end{aligned}
\end{equation}
We term (\ref{eq:flowModel1}) an ``Oscillation + Gradient Flow" because the equation
\begin{align}
\dot{\xb} &=  - (\Lb\otimes \Ib_m) \vb  \notag \\
\dot{\vb} &= (\Lb\otimes \Ib_m) \xb \notag
\end{align}
yields oscillating trajectories for $\xb(t)$, while $\dot{\xb}=-\nabla U(\xb)$ is a gradient flow.

Note that in the flow (\ref{eq:flowModel1}), the state variable $[\xb^\top_i(t) \ \vb^\top_i(t)]^\top$ of node $i$ obeys the evolution
\begin{align}
\dot{\xb}_i(t) &= -\sum_{j\in\mathcal{N}_i}(\vb_i(t)-\vb_j(t)) - (\hb_i\hb_i^\top\xb_i(t)-z_i\hb_i) \notag \\
\dot{\vb}_i(t) &= \sum_{j\in\mathcal{N}_i}(\xb_i(t)-\xb_j(t)) \notag
\end{align}
Therefore, besides the equation $\hb_i^\top\yb=z_i$ that node $i$ possesses, it only needs to communicate with its neighbors to obtain their states in order to implement (\ref{eq:flowModel1}). The flow (\ref{eq:flowModel1}) is {\it distributed} in that sense.

\subsection{Discussions}
\subsubsection{Relation to A-H-U Flow}
Consider a constrained optimization problem as
\begin{equation} \label{eq:wang_opt}
\begin{aligned}
\min_{\xb}\qquad &  f(\xb) \\
{\rm s.t.} \qquad  &   \mathbf{F}\xb=\bb
\end{aligned}
\end{equation}
where $f(\cdot):\R^n\to\R$ is a differentiable function, $\mathbf{F}\in\R^{m\times n}$ and $\bb\in\R^m$. The well-known Arrow-Hurwicz-Uzawa (A-H-U) flow introduced in \cite{arrow:58}  provides under appropriate conditions a continuous-time solver defined by
\begin{equation}
\begin{aligned}
\dot{\xb} &= -\nabla_{\xb}f(\xb)-\mathbf{F}^\top \vb \\
\dot{\vb} &= \mathbf{F}\xb - \bb. \label{eq:wang_model}
\end{aligned}
\end{equation}
In particular, if $f$ is strictly convex and $\mathbf{F}$ has full row rank, then (see \cite{arrow:58}\cite{wang:elia:11}) along the flow (\ref{eq:wang_model}), $\xb(t)$ will converge to the optimal point of (\ref{eq:wang_opt}) and $\vb(t)$ will converge to a Lagrangian multiplier of (\ref{eq:wang_opt}).

As one can see, the flow (\ref{eq:flowModel1}) is a form of the A-H-U flow (\ref{eq:wang_model}) with the cost function $f(\xb)$ being the given $U(\xb)$ and the constraint $\mathbf{F}\xb=\bb$ given by $(\Lb\otimes\Ib_m)\xb=0$. However, neither the Laplacian $\Lb$ is a full-rank matrix, nor is $U(\xb)$ strictly convex. Therefore, the sufficiency results and analysis for the A-H-U flow established in \cite{wang:elia:11} cannot be applied directly to the flow (\ref{eq:flowModel1}).
\subsubsection{Relation to Wang-Elia/Gharesifard-Cort\'{e}s Flows}
In \cite{wang:elia:10} and \cite{cortez:14}, an alternative distributed flow for solving
\begin{equation} \label{eq:opdistributed}
\begin{aligned}
\min_{\xb=(\xb_1^\top \dots \xb_N^\top)^\top}\qquad & \sum_{i=1}^N f_i(\xb_i)  \\
{\rm s.t.} \qquad  &   \xb_i\in \mathbb{R}^m, i=1,\dots, N \\
& \xb_1=\dots=\xb_N
\end{aligned}
\end{equation}
was proposed, where each $f_i:\mathbb{R}^m \mapsto \mathbb{R}$ is a locally Lipschitz and convex function known to node $i$ only. Viewing the component $|\hb_i^\top\xb_i-z_i|^2$ in $U(\xb_1,\dots,\xb_N)$ as the $f_i$ in (\ref{eq:opdistributed}),  the Wang-Elia/Gharesifard-Cort\'{e}s flow  \cite{wang:elia:10,cortez:14} reads as
\begin{equation}
\begin{aligned}
\dot{\xb}  &=-{\alpha} (\Lb\otimes\Ib_m){\xb}-   (\Lb\otimes\Ib_m){\vb} - \tilde{\Hb}\xb {+} \zb_H \\
\dot{\vb} &= (\Lb\otimes\Ib_m)\xb. \label{eq:wang-elia-cortes}
\end{aligned}
\end{equation}
where $\alpha$ is a positive number, $\zb_H=[z_1\hb_1^\top \ \dots \ z_H\hb_N^\top]^\top$ and $\tilde{\Hb}=\diag(\hb_1\hb_1^\top, \dots, \hb_N\hb_N^\top)$
since by direct calculation we know $
\nabla U(\xb) = \tilde{\Hb}\xb-\zb_H
$. With fixed interaction graph and suitable choice of $\alpha$, the flow (\ref{eq:wang-elia-cortes}) is a least squares solver for (\ref{eq:linear_equation})  even for balanced directed networks (Theorem 5.4, \cite{cortez:14}). The system (\ref{eq:flowModel1}) under consideration in the current work certainly has a  simpler structure compared to (\ref{eq:wang-elia-cortes}), which is desirable in large-scale applications. Moreover, it is also useful to establish a clear understanding of the convergence conditions of the system (\ref{eq:flowModel1}) under general conditions (which are currently missing in the literature),the more so considering the historical context.

\section{Continuous Flow} \label{sec:3}
In this section, we study the behavior of the flow (\ref{eq:flowModel1}) in terms of convergence to a least squares solution for the $\xb_i(t)$, and present necessary and sufficient conditions for convergence.

\subsection{Convergence Result}
We present the following result.
\begin{thm}\label{thm:1}
Assume that $N>m$ and $\rank(\Hb)=m$. Let $\yb^\ast=(\Hb^\top \Hb)^{-1}\Hb^\top\zb$ be the unique least squares solution of (\ref{eq:linear_equation}). Define $\mathcal{S}_{\Lb}$ as the set of all complex eigenvectors of $\Lb$ and
for $\alphab\in\mathcal{S}_{\Lb}$ with $\alphab[i]$ denoting the $i$-th entry,
\[
\mathcal{I}_{\alphab} := \{i:\alphab[i]\neq 0, \ \alphab=\big[\alphab[1] \ \alphab[2] \ \dots \ \alphab[N]\big]^\top\}.
\]
Then:
\begin{enumerate}[(i)]
\item If $\spa\{\hb_i:i\in\mathcal{I}_{\alphab}\}=\R^m$ for all $\alphab\in\mathcal{S}_{\Lb}$, there holds
\[
\lim_{t\to\infty}\xb_i(t) = \yb^\ast, i=1,\dots,N
\]
along the flow (\ref{eq:flowModel1}). Further $\vb(t)$ along (\ref{eq:flowModel1}) converges to a Lagrange multiplier associated with a solution of the optimization problem
\begin{equation} \label{eq:optproblem_cont}
\begin{aligned}
\min_{\xb}\qquad &  U(\xb) \\
{\rm s.t.} \qquad  &   (\Lb\otimes\Ib_m)\xb=0
\end{aligned}
\end{equation}
\item If there exists $\alphab\in\mathcal{S}_{\Lb}$ such that $\dim(\spa\{\hb_i:i\in\mathcal{I}_{\alphab}\})<m$, then there exist trajectories of $\xb(t)$ along (\ref{eq:flowModel1}) which do not converge.
\end{enumerate}
\end{thm}
\noindent{\em Proof.}
Recall that $\nabla U(\xb) = \tilde{\Hb}\xb-\zb_H$ where $\tilde{\Hb}$ is the nonnegative matrix $\diag(\hb_1\hb_1^\top, \dots, \hb_N\hb_N^\top)$ and $\zb_H=[z_1\hb_1^\top \ \dots \ z_H\hb_N^\top]^\top$. We rewrite (\ref{eq:flowModel1}) as
\begin{equation}
\begin{aligned}
\dot{\xb} &=  - (\Lb\otimes \Ib_m) \vb - \tilde{\Hb}\xb + \zb_H \\
\dot{\vb} &= (\Lb\otimes \Ib_m) \xb. \label{eq:flowModel2}
\end{aligned}
\end{equation}
Suppose there exists an equilibrium $(\xb^\ast, \vb^\ast)$ of (\ref{eq:flowModel2}), i.e.
\begin{equation} \label{eq:equilib}
\begin{aligned}
0 &=  - (\Lb\otimes \Ib_m) \vb^\ast - \tilde{\Hb}\xb^\ast + \zb_H  \\
0 &= (\Lb\otimes \Ib_m) \xb^\ast.
\end{aligned}
\end{equation}

It is worth noting that (\ref{eq:equilib}) specifies exactly the Karush-Kuhn-Tucker conditions on $(\xb^\ast, \vb^\ast)$ for the optimization problem (\ref{eq:optproblem_cont}) \cite{bertsekas1999nonlinear}. Since $U(x)$ is a convex function and the constraints in (\ref{eq:optproblem_cont}) are equality constraints, Slater's condition holds \cite{boyd2004convex}. Therefore the optimal points of the primal problem and dual problem are the same, i.e., $\xb^\ast$ is an optimal solution to (\ref{eq:optproblem_cont}) and any optimal solution of (\ref{eq:optproblem_cont}) must have the form
\[
\mathbf{1}\otimes\yb^\ast,
\]
where $\yb^\ast$ is a least squares solution to (\ref{eq:linear_equation}). We know $\yb^\ast$ is unique because $\rank(\Hb)=m$. Since $\xb^\ast=1\otimes\yb^\ast$, then $\xb^\ast$ is also unique.
Note however that $\vb^\ast$ is not necessarily unique. Define the variables $\hat{\xb} = \xb - \xb^\ast$, $\hat{\vb} = \vb-\vb^\ast$.
Then
\begin{equation}
\begin{aligned}
\dot{\hat{\xb}} &= \dot{\xb} - \dot{\xb}^\ast
= - (\Lb\otimes \Ib_m) \hat{\vb} - \tilde{\Hb}\hat{\xb} \\
\dot{\hat{\vb}} &= (\Lb\otimes \Ib_m) \hat{\xb} .
\label{eq:dynsys}
\end{aligned}
\end{equation}
Denote $\hat{\ub}(t) = [{\hat{\xb}}(t)^\top \ {\hat{\vb}}(t)^\top]^\top$ and
\[
\Mb =
\begin{bmatrix}
-\tilde{\Hb} & -\Lb\otimes\Ib_m \\
\Lb\otimes\Ib_m & 0
\end{bmatrix}.
\]
Then (\ref{eq:dynsys}) is a linear system with the form $\dot{\hat{\ub}}=\Mb\hat{\ub}$.
Consider the following Lyapunov function:
\[
V(\hat{\xb},\hat{\vb})
=\frac{1}{2}\|\hat{\ub}\|^2
=\frac{1}{2}(\|\hat{\xb}\|^2+\|\hat{\vb}\|^2).
\]
Since
\begin{equation} \label{eq:Lya_Derivative}
\begin{aligned}
\dot{V}  &=  - \hat{\xb}^\top(\Lb\otimes \Ib_m) \hat{\vb} - \hat{\xb}^\top\tilde{\Hb}\xb
+\hat{\vb}^\top(\Lb\otimes \Ib_m) \hat{\xb} \\
&= - \hat{\xb}^\top\tilde{\Hb} \hat{\xb} \le 0,
\end{aligned}
\end{equation}
$\hat{\ub}(t)$ is bounded for any finite initial values $\hat{\xb}(0)$, $\hat{\vb}(0)$, namely $\hat{\ub}(0)$. Therefore, we conclude:

{\bf C1}. $\Re(\lambda) \le 0$ for all $\lambda\in\sigma(\Mb)$.

{\bf C2}. If $\Re(\lambda)=0$, then $\lambda$ has equal algebraic and geometric multiplicity.

\noindent\emph{(\lowercase\expandafter{\romannumeral 1})}. Suppose $\spa\{\hb_i:i\in\mathcal{I}_{\alphab}\}=\R^m$ for all $\alphab\in\mathcal{S}_{\Lb}$. We proceed to prove the convergence of $\hat{\xb}(t)$ and $\hat{\vb}(t)$. The proof contains two steps.

\noindent Step 1. We prove $\Mb$ does not have a purely imaginary eigenvalue if $\spa\{\hb_i:i\in\mathcal{I}_{\alphab}\}=\R^m$ for all $\alphab\in\mathcal{S}_{\Lb}$, using a contradiction argument. Thus suppose $\lambda=\imath r\neq 0$ where $r\in\R$ is an eigenvalue of $\Mb$ with a corresponding eigenvector $\betab=[\betab_a^\top \ \betab_b^\top]\in\C^{2Nm}$, where $\betab_a\in\C^{Nm}$,  $\betab_b\in\C^{Nm}$. Let $\hat{\ub}(0)=\betab$. Then
\[
\hat{\ub}(t) = e^{\Mb t}\hat{\ub}(0) = e^{\imath rt}\hat{\ub}(0).
\]
Therefore, $\|\hat{\ub}(t)\|^2=\|\hat{\ub}(0)\|^2$ for all $t$.

On the other hand, according to (\ref{eq:Lya_Derivative}),
\begin{equation}
\begin{aligned}
\frac{d}{dt}(\frac{1}{2}\|\hat{\ub}(t)\|^2)
&= -\hat{\xb}^\top(t)\tilde{\Hb}\hat{\xb}(t) \notag \\
&= -\hat{\xb}^\top(0)e^{\imath rt}\tilde{\Hb}e^{\imath rt}\hat{\xb}(0) \notag \\
&= -e^{\imath 2rt}\betab_a^\top\tilde{\Hb}\betab_a.
\end{aligned}
\end{equation}
Consequently, there must hold $\tilde{\Hb}\betab_a=0$.
Next, based on
\[
\begin{bmatrix}
-\tilde{\Hb} & -\Lb\otimes\Ib_m \\
\Lb\otimes\Ib_m & 0
\end{bmatrix}
\begin{bmatrix}
\betab_a \\
\betab_b \\
\end{bmatrix}
= \imath r
\begin{bmatrix}
\betab_a \\
\betab_b \\
\end{bmatrix},
\]
we know
\begin{equation}\label{eq:beta_a_beta_b}
\begin{aligned}
-(\Lb\otimes\Ib_m)\betab_b &= \imath r\betab_a \\
(\Lb\otimes\Ib_m)\betab_a &= \imath r\betab_b.
\end{aligned}
\end{equation}
Since ${ \betab}\neq 0$, neither of ${\betab_a}$ nor ${\betab_b}$ can be zero. By simple calculation, we have
\begin{equation} \label{eq:Lsquare}
\begin{aligned}
(\Lb\otimes\Ib_m)^2\betab_a &= r^2\betab_a \\
(\Lb\otimes\Ib_m)^2\betab_b &= r^2\betab_b,
\end{aligned}
\end{equation}
i.e., $\betab_a$ and $\betab_b$ are both eigenvectors of $(\Lb\otimes\Ib_m)^2$ corresponding to $r^2$. From (\ref{eq:Lsquare}), we know
\begin{equation}
\begin{aligned}
(\Lb^2\otimes\Ib_m)\betab_a &= r^2\betab_a. \notag
\end{aligned}
\end{equation}
Based on the properties for eigenvectors of the Kronecker product of two matrices (Theorem 13.12 \cite{alan:kroneckerproduct}), we know there exist $(r^2,\alphab_a)$ and $\etab_a$ such that $\Lb^2\alphab_a=r^2\alphab_a$ and $\betab_a=\alphab_a\otimes\etab_a$ with $\alphab_a\in\C^N$ and $\etab_a\in\C^m$. It is trivial that if $\Lb^2\alphab_a=r^2\alphab_a$, $\Lb\alphab_a=|r|\alphab_a$, i.e., $\alphab_a$ is an eigenvector of $\Lb$ corresponding to eigenvalue $|r|$.
Denote
\[
\betab_a =
\begin{bmatrix}
\betab_a^{[1]} \\
\betab_a^{[2]} \\
\vdots \\
\betab_a^{[N]}
\end{bmatrix}, \
\betab_a^{[i]}\in\C^m,\ i=1,2,\dots,N
\]
and
\[
\etab_a =
\begin{bmatrix}
\etab_a[1] \\
\etab_a[2] \\
\vdots \\
\etab_a[N]
\end{bmatrix}, \ \etab_a[i]\in\C, \ i=1,2,\dots,N.
\]
It is apparent that $\betab_a^{[i]} = \alphab_a[i]\etab_a$ if $i\in\mathcal{I}_{\alphab_a}$ and $\betab_a^{[i]}=0$ otherwise. Then noting that
\[
\tilde{\Hb}\betab_a=
\begin{bmatrix}
\alphab_a[1]\hb_1\hb_1^\top\etab_a \\
\alphab_a[2]\hb_2\hb_2^\top\etab_a \\
\vdots \\
\alphab_a[N]\hb_N\hb_N^\top\etab_a \\
\end{bmatrix}
=0,
\]
we get $\alphab_a[i]\hb_i\hb_i^\top\etab_a=0$ for $i=1,2,\dots,N$, which implies that
\begin{equation} \label{eq:eta}
\hb_i^\top\etab_a=0, \ i\in\mathcal{I}_{\alphab_a}.
\end{equation}
Because $\spa\{\hb_i:i\in\mathcal{I}_{\alphab_a}\}=\R^m$, there must hold $\etab_a=0$. In turn, $\betab_a$ must be zero, leading to $\betab_b=0$ with (\ref{eq:beta_a_beta_b}). Therefore $\Mb$ does not have purely imaginary eigenvalues.

Based on ${\bf C1}$, ${\bf C2}$ and the fact that $\Mb$ has no purely imaginary eigenvalue, it follows that $\hat{\xb}(t)$ and $\hat{\vb}(t)$ converge.

\noindent Step 2. In this step, we establish the limits of $\hat{\xb}(t)$ and $\hat{\vb}(t)$ by studying the zero eigenspace of $\Mb$,  thereby obtaining the convergence property for $\xb(t)$ and $\vb(t)$. Suppose $\deltab=[\deltab_a^\top \ \deltab_b^\top]^\top$ is one of the eigenvectors of $\Mb$ corresponding to zero eigenvalue with $\deltab\in\R^{2Nm}$ and $\deltab_a$, $\deltab_b\in\R^{Nm}$, i.e., $\Mb \deltab = 0$.
Consider a solution $\hat{\ub}(t)$ of \eqref{eq:dynsys} with
$\hat{\ub}(0)=\deltab$. We see from the derivative of the Lyapunov function and $\Mb\deltab=0$ that
\begin{equation}
\begin{aligned}
\tilde{\Hb}\deltab_a &= 0 \notag \\
(\Lb\otimes\Ib_m)\deltab_a &= 0 \notag \\
(\Lb\otimes\Ib_m)\deltab_b &= 0 \notag
\end{aligned}
\end{equation}
Then there exist $\etab_a\in\R^m$ and $\etab_b\in\R^m$ such that $\deltab_a=\mathbf{1}\otimes\etab_a$ and $\deltab_b=\mathbf{1}\otimes\etab_b$. Since $\tilde{\Hb}\deltab_a=0$ and $\rank(\tilde{\Hb})=m$, $\deltab_a=0$, i.e., $\deltab$ must be in the form $\deltab = [0 \ \deltab_b^\top]^\top$ with $\deltab_b=\mathbf{1}\otimes\etab_b$. Note that the algebraic and geometric multiplicity of the zero eigenvalue of $\Mb$ is $m$. Now we decompose $\Mb$ into its Jordan canonical form $\Mb=\mathbf{T}\mathbf{J}\mathbf{T}^{-1}$:
\[
\begin{aligned}
\mathbf{T} &= [\deltab_1 \ \deltab_2 \ \cdots \deltab_{m} \ \cdots], \\
\mathbf{T}^{-1} &= [\deltab_1' \ \deltab_2' \ \cdots \ \deltab_{m}' \ \cdots]^\top
\end{aligned}
\]
where $\deltab_i$ and $\deltab_i'^\top$ with $i=1,2,\dots,m$ are mutually orthogonal right and left eigenvectors respectively of $\Mb$ all corresponding to zero eigenvalues and all with the form of $\deltab_i=[0 \ \deltab_{ib}^\top]^\top$ and $\deltab_i'^\top=[0 \ \deltab_{ib}'^\top]$. Then
\[
\lim_{t\to\infty}\hat{\ub}(t) = \sum_{i=1}^{m}\deltab_i\deltab_i'^\top\hat{\ub}(0),
\]
which implies that
\[
\begin{aligned}
\lim_{t\to\infty}\hat{\xb}(t) &= 0, \\
\lim_{t\to\infty}\hat{\vb}(t) &= \mathbf{W}_{\mathcal{G}}\hat{\vb}(0)
\end{aligned}
\]
with $\mathbf{W}_{\mathcal{G}}=\sum_{i=1}^{m}\deltab_{ib}\deltab_{ib}'^\top$.
Thus we can conclude that $\xb(t)$ converges to $\xb^\ast=1\otimes\yb^\ast$ while $\vb(t)$ converges to a constant associated with the initial value $\vb(0)$, which can be obtained by
\begin{equation}\notag
\begin{aligned}
\lim\limits_{t\to\infty}\vb(t)
&= \vb^\ast + \lim\limits_{t\to\infty}\hat{\vb}(t) \\
&= \vb^\ast + \mathbf{W}\hat{\vb}(0) \\
&= \vb^\ast + \mathbf{W}(\vb(0)-\vb^\ast) \\
&= (\Ib_{Nm}-\mathbf{W})\vb^\ast + \mathbf{W}\vb(0).
\end{aligned}
\end{equation}
Recall that $\vb^\ast$ was chosen satisfying (\ref{eq:equilib}).
This completes the proof of (\lowercase\expandafter{\romannumeral 1}).

\noindent\emph{(\lowercase\expandafter{\romannumeral 2})}. Suppose there exists $\alphab_a\in\mathcal{S}_{\Lb}$ with $\Lb\alphab_a=r\alphab_a$ such that $\dim(\spa\{\hb_i:i\in\mathcal{I}_{\alphab_a}\})<m$. Then there must exist $\etab_a\neq 0$ satisfying that
\[
\hb_i^\top\etab_a=0, \ i\in\mathcal{I}_{\alphab_a}
\] 	
Let $\betab=[\betab_a^\top \ \betab_b^\top]^\top$ with $\betab_a=\alphab_a\otimes\etab_a$ and $\betab_b=\frac{(\Lb\otimes\Ib_m)\betab_a}{\imath r}$. It is easy to check that
\begin{equation}
\begin{aligned}
\Mb\betab &=
\begin{bmatrix}
-\tilde{\Hb} & -\Lb\otimes\Ib_m \\
\Lb\otimes\Ib_m & 0
\end{bmatrix}
\begin{bmatrix}
\betab_a \\
\betab_b
\end{bmatrix}
\\
&=
\begin{bmatrix}
-(\Lb\otimes\Ib_m)\betab_b \\
(\Lb\otimes\Ib_m)\betab_a
\end{bmatrix}
= \imath r
\begin{bmatrix}
\betab_a \\
\betab_b
\end{bmatrix}.
\end{aligned}
\end{equation}
Therefore, $\Mb$ has a purely imaginary eigenvalue. Hence, $\xb(t)$ and $\vb(t)$ do not converge for generic initial conditions.

We have now completed the proof of Theorem \ref{thm:1}.
\hfill$\square$

\subsection{Generic Feasibility of $\Hb$}
\begin{prop}
Let $\Lb$ be the Laplacian of a graph $\mathcal{G}$ with $|\mathcal{I}_{\alphab}|\ge m$ for all $\alphab\in\mathcal{S}_{\Lb}$. Then the convergence condition of Theorem \ref{thm:1}, viz. $\spa\{\hb_i:i\in\mathcal{I}_{\alphab}\}=\R^m$, is satisfied for generic $\Hb\in\R^{N\times m}$, i.e., there does not exist an open nonempty subset in $\R^{N\times m}$ of $\Hb$, for which the convergence condition in Theorem \ref{thm:1} is not satisfied.
\end{prop}
\noindent{\em Proof.}
Let
\[
\Hb_{i_1,\dots,i_m} = \begin{bmatrix}
\hb_{i_1} \\
\vdots \\
\hb_{i_m}
\end{bmatrix}, \
1\le i_1<\dots<i_m \le N.
\]
Introduce
\begin{gather}
\mathcal{W} = \bigcup\limits_{\substack{i_1,\dots,i_m:\\ 1\le i_1<\dots<i_m \le N}} \mathcal{W}_{i_1,\dots,i_m} \notag
\end{gather}
where
\[
\mathcal{W}_{i_1,\dots,i_m} = \{ \Hb\in\R^{N\times m}:\ \det(\Hb_{i_1,\dots,i_m})= 0\}.
\]
It can be noted that $\mathcal{W}$ is the set of $\Hb$ for which the convergence condition in Theorem \ref{thm:1} is not satisfied. Since $\mathcal{W}_{i_1,\dots,i_m}\neq\R^{N\times m}$, by identity theorem for holomorphic functions \cite{ablowitz2003complex}, $\mathcal{W}_{i_1,\dots,i_m}$ does not contain a nonempty open subset of $\R^{N\times m}$, i.e., $\mathcal{W}$ does not contain a nonempty open subset of $\R^{N\times m}$. This completes the proof. \hfill$\square$

\subsection{Graph Feasibility}
In this section, we consider several fundamental graphs to investigate the feasibility of the convergence condition presented in Theorem \ref{thm:1}. Suppose $N>2$. For a number of graphs we will first determine the minimum value of $|\mathcal{I}_{\alphab}|$.  The collection of values and the implications for solvability of the least squares problem will  be interpreted for all the graphs at the end of the calculations.

\noindent{\bf [Path Graph]} It is known from \cite{fuhrmann2015mathematics} that all the eigenvalues of its Laplacian $\Lb$ are distinct with eigenvectors in the set of
$
\mathcal{S}_{\Lb} =
\{
\alphab_k: \
\alphab_k[v] = \cos\frac{(k-1)(2v-1)\pi}{2N}, \ v=1,\dots,N;
\ k=1,\dots,N
\}
$.
We discuss two cases:
\begin{enumerate}[(i)]
\item Let $N=2^l, \ l=2,3,4,\dots$. Then it is obvious that there do not exist $v$ and $k$ such that $\alphab_k[v]=0$. Therefore $|\mathcal{I}_{\alphab}|=N$ for all $\alphab$.
\item Let $N=3l, \ l=1,2,3,\dots$. Then any $\alphab_k\in\mathcal{S}_{\Lb}$ contains at most $l$ zero entries. Therefore $\min\limits_{\alphab\in\mathcal{S}_{\Lb}}|\mathcal{I}_{\alphab}|=\frac{2}{3}N$.
\end{enumerate}

\noindent{\bf [Ring Graph]} We know from \cite{fuhrmann2015mathematics} that if $N$ is odd, then zero is the only eigenvalue of multiplicity one with eigenvector $[1 \ 1 \ \dots \ 1]^\top$, while all the other eigenvalues have multiplicity two with a basis of two orthogonal eigenvectors
\begin{equation}\label{eq:ringGraph}
\begin{bmatrix}
1 \\
\cos\frac{2k\pi}{N} \\
\cos\frac{4k\pi}{N} \\
\vdots \\
\cos\frac{2(N-1)k\pi}{N}
\end{bmatrix}
, \
\begin{bmatrix}
0 \\
\sin\frac{2k\pi}{N} \\
\sin\frac{4k\pi}{N} \\
\vdots \\
\sin\frac{2(N-1)k\pi}{N}
\end{bmatrix}
\end{equation}
with $k = 1,\dots,N-1$. If $N$ is even, then zero and the largest eigenvalue are the only two eigenvalues of multiplicity one with eigenvectors $[1 \ 1 \ \dots \ 1]^\top$ and $[1 \ -1 \ 1 \ \dots \ 1]^\top$ respectively, while all the other eigenvalues have multiplicity two with a basis of two orthogonal eigenvectors with the same form (\ref{eq:ringGraph}) and $k = 1,\dots,N-1,\ k\neq \frac{N}{2}$. Note that the eigenspaces of $k=p$ and $k=q$ are the same if and only if $p+q=N$ and $1\le p,q\le N$.
\begin{enumerate}[(i)]
\item If $N$ is a prime number, then any $\alphab\in\mathcal{S}_{\Lb}$ contains at most one zero entry. Therefore $\min\limits_{\alphab\in\mathcal{S}_{\Lb}}|\mathcal{I}_{\alphab}|=N-1$.
\item If $N=3l, \ l=1,2,3,\dots$, then any $\alphab\in\mathcal{S}_{\Lb}$ contains at most $l$ zero entries. Therefore $\min\limits_{\alphab\in\mathcal{S}_{\Lb}}|\mathcal{I}_{\alphab}|=\frac{2}{3}N$.
\item If $N=2^l, \ l=3,4,\dots$, then any $\alphab\in\mathcal{S}_{\Lb}$ contains at most $2^{l-1}$ zero entries. Therefore $\min\limits_{\alphab\in\mathcal{S}_{\Lb}}|\mathcal{I}_{\alphab}|=\frac{1}{2}N$.
\end{enumerate}

\noindent{\bf [Star Graph]} We know that its Laplacian has an eigenvalue zero of multiplicity one with eigenvector $\alphab_1=[1 \ \dots \ 1]^\top$, an eigenvalue $N$ of multiplicity one with eigenvector $\alphab_N=[1-N \ 1 \ \dots \ 1]^\top$ and eigenvalue one with multiplicity $N-2$ and a set of associated eigenvectors
$
\{\alphab_k|\mathbf{1}^\top\alphab_k=0,\ \alphab_k\neq p[1-N \ 1 \ \dots \ 1]^\top,\ p\in\R; \ k=2,3,\dots,N-1\}
$.
Thus $\alphab_k$ has at most $N-2$ zero entries. Therefore $\min\limits_{\alphab\in\mathcal{S}_{\Lb}}|\mathcal{I}_{\alphab}|=2$.

\noindent{\bf [Complete Graph]} It is known from \cite{completegraph2009} that its Laplacian has an eigenvalue zero of multiplicity one with eigenvector $\alphab_1=[1 \ \dots \ 1]^\top$ and eigenvalue $N$ with multiplicity $N-1$ and a set of associated eigenvectors $\{\alphab_k|\mathbf{1}^\top\alphab_k=0; \ k=2,3,\dots,N\}$. Then it can be concluded that $\alphab_k$ has at most $N-2$ zero entries. Therefore $\min\limits_{\alphab\in\mathcal{S}_{\Lb}}|\mathcal{I}_{\alphab}|=2$.


For star and complete graphs, there holds that $\min\limits_{\alphab\in\mathcal{S}_{\Lb}}|\mathcal{I}_{\alphab}|=2$. This means that as long as $m>2$, the sufficient convergence condition in Theorem \ref{thm:1} will not hold. On the other hand, for path and ring graphs,
\[
\min\limits_{\alphab\in\mathcal{S}_{\Lb}}|\mathcal{I}_{\alphab}| \approx \mathcal{O}(N).
\]
Therefore, if $N\gg m$, it is relatively easy for the sufficient condition in Theorem \ref{thm:1} to hold.

\subsection{Numerical Examples}
We now provide several numerical examples illustrating the result of Theorem \ref{thm:1}.

\begin{figure}
\centering
\includegraphics[width=1.6in]{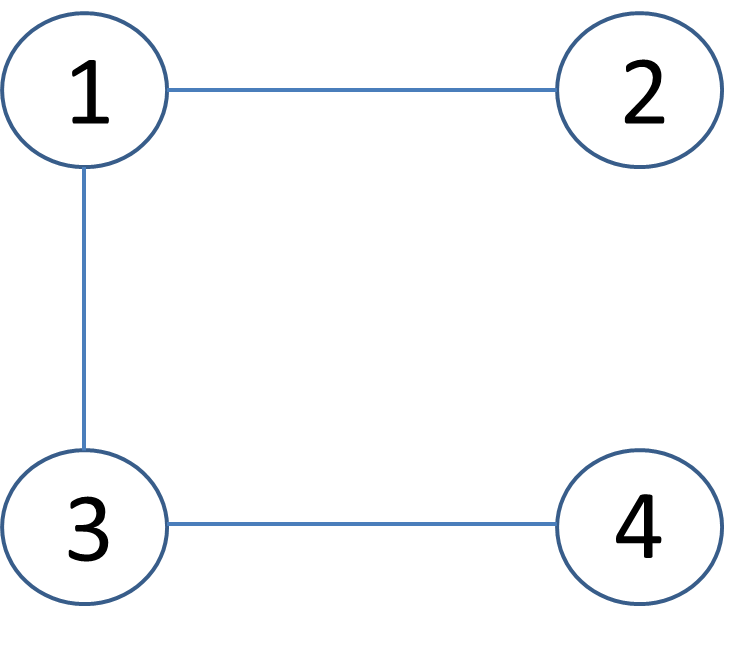}
\caption{A constant, undirected and connected graph with four nodes considered in Example 1 and Example 3.}
\label{fig:conv_graph}
\end{figure}

\noindent{\bf Example 1}
Consider a linear equation in the form of (\ref{eq:linear_equation}) where $\yb\in\R^2$, and
\[
\Hb =
\begin{bmatrix}
0 & 1\\
3 & 0\\
2 & 0\\
1 & 0
\end{bmatrix},
\quad
\zb =
\begin{bmatrix}
-1\\
0\\
-2\\
2
\end{bmatrix}.
\]
Then the equation has a unique least squares solution $\yb^\ast=[-0.1429 \ -1]^\top$. Let the interaction graph $\mathcal{G}=(\mathcal{V},\mathcal{E})$ be given in Figure \ref{fig:conv_graph}. It can be noted that graph $\mathcal{G}$ is a path graph that satisfies Case [{\bf Path Graph}](\lowercase\expandafter{\romannumeral 1}) in Section 3.3. It is verified that there is no zero entry in any eigenvectors of $\Lb$ by calculating four eigenvectors as $[-0.5 \ -0.5 \ -0.5 \ -0.5]^\top$, $[0.27 \ 0.65 \ -0.27 \ -0.65]^\top$, $[-0.5 \ 0.5 \ -0.5 \ 0.5]^\top$ and $[0.65 \ -0.27 \ -0.65 \ 0.27]^\top$ corresponding to the eigenvalues $0$, $0.59$, $2$ and $3.41$, respectively. One can easily verify the condition of Theorem \ref{thm:1}.(\lowercase\expandafter{\romannumeral 1}) holds. Let $\xb(0)=[-2 \ -0.5 \ -1.8 \ -1.5 \ 1.8 \ -0.6 \ 1.9 \ -1.4]^\top$, $\vb(0)=0$ be the initial condition.

\begin{figure}
\centering
\includegraphics[width=3in]{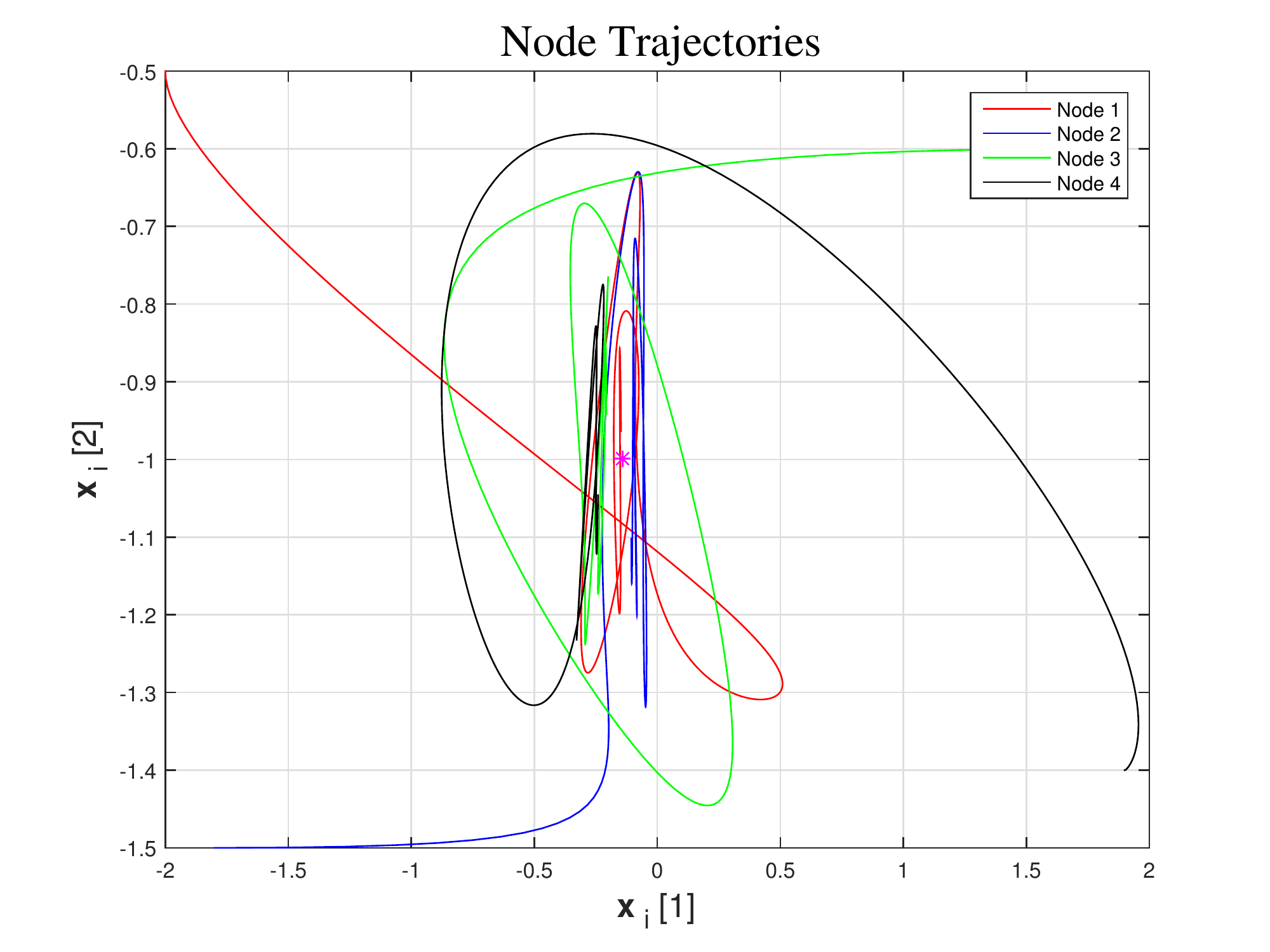}
\caption{The trajectories of $\xb$ over the graph in Figure \ref{fig:conv_graph}. We can see that if the initial condition is $[-2 \ -0.5 \ -1.8 \ -1.5 \ 1.8 \ -0.6 \ 1.9 \ -1.4]^\top$, it will converge to $[-0.1429 \ -0.1 \ -0.1429 \ -0.1 \ -0.1429 \ -0.1 \ -0.1429 \ -0.1]^\top$.}
\label{fig:conv_xx}
\end{figure}

\begin{figure}
\centering
\includegraphics[width=3in]{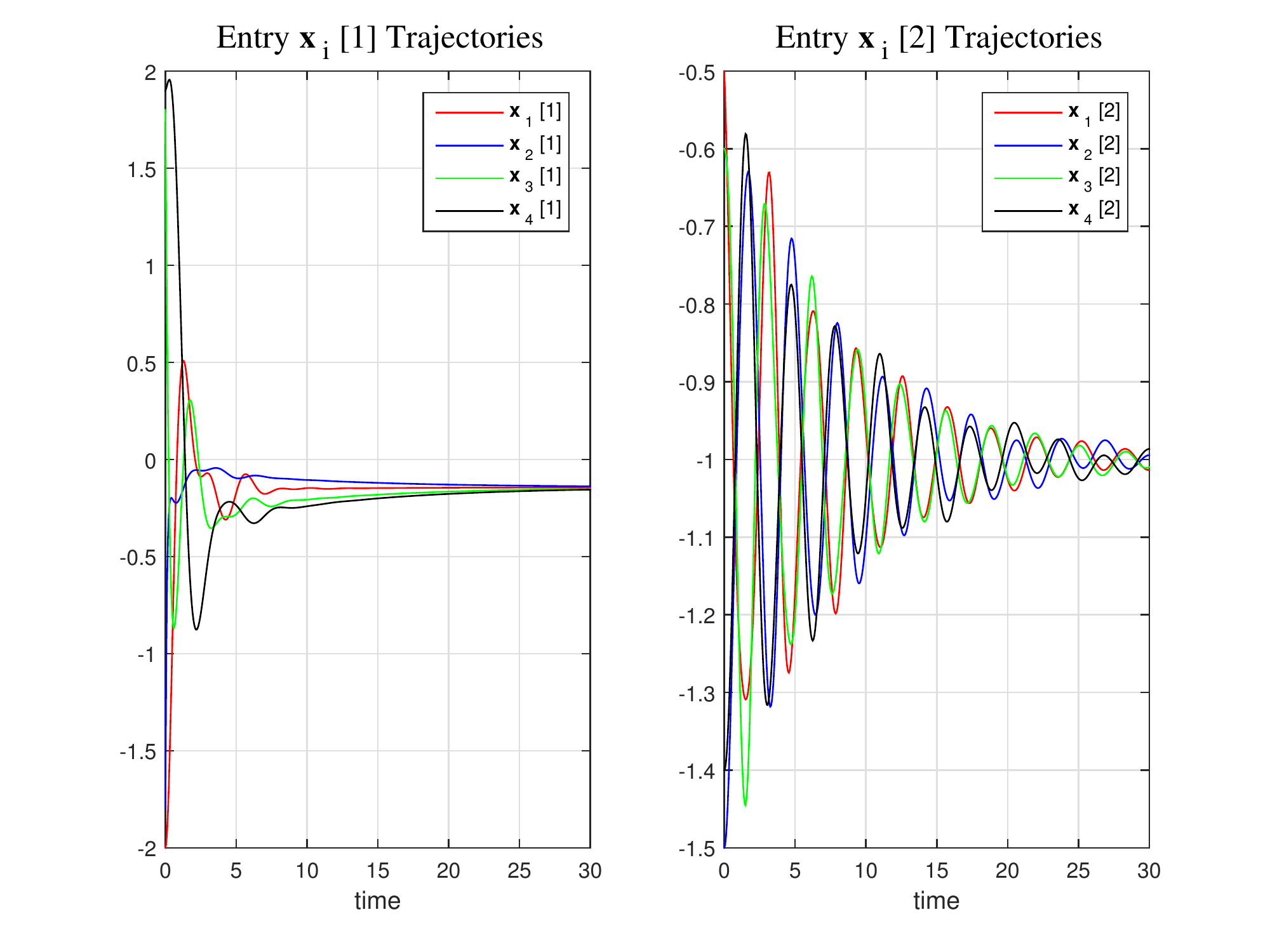}
\caption{The trajectories of $\xb_i[1]$ and $\xb_i[2]$ for $i=1,2,3,4$ over the graph in Figure \ref{fig:conv_graph}, which converge to $-0.1429$ and $-1$ respectively.}
\label{fig:conv}
\end{figure}

In Figure \ref{fig:conv_xx} we first plot the trajectories of the $\xb_i(t), i=1,2,3,4$ in $\R^2$. As we can see, all $\xb_i(t)$ converge to $\yb^\ast$. A further confirmation from the two entries $(\xb_i(t))[1]$ and $(\xb_i(t))[2]$ is shown in Figure \ref{fig:conv}. The numerical result is consistent with Theorem 1.(\lowercase\expandafter{\romannumeral 1}).

\begin{figure}
\centering
\includegraphics[width=1.4in]{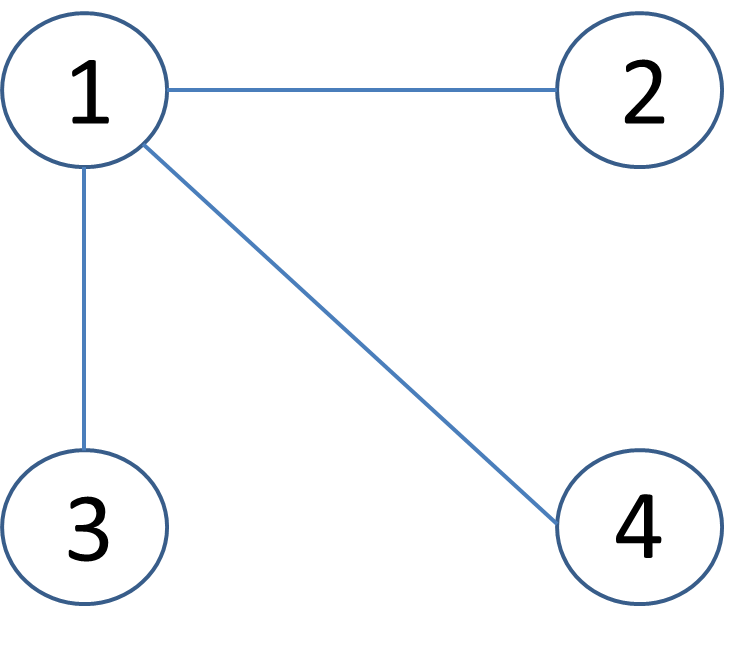}
\caption{A constant, undirected and connected graph with four nodes considered in Example 2 and Example 4.}
\label{fig:div_graph}
\end{figure}

\begin{figure}
\centering
\includegraphics[width=3in]{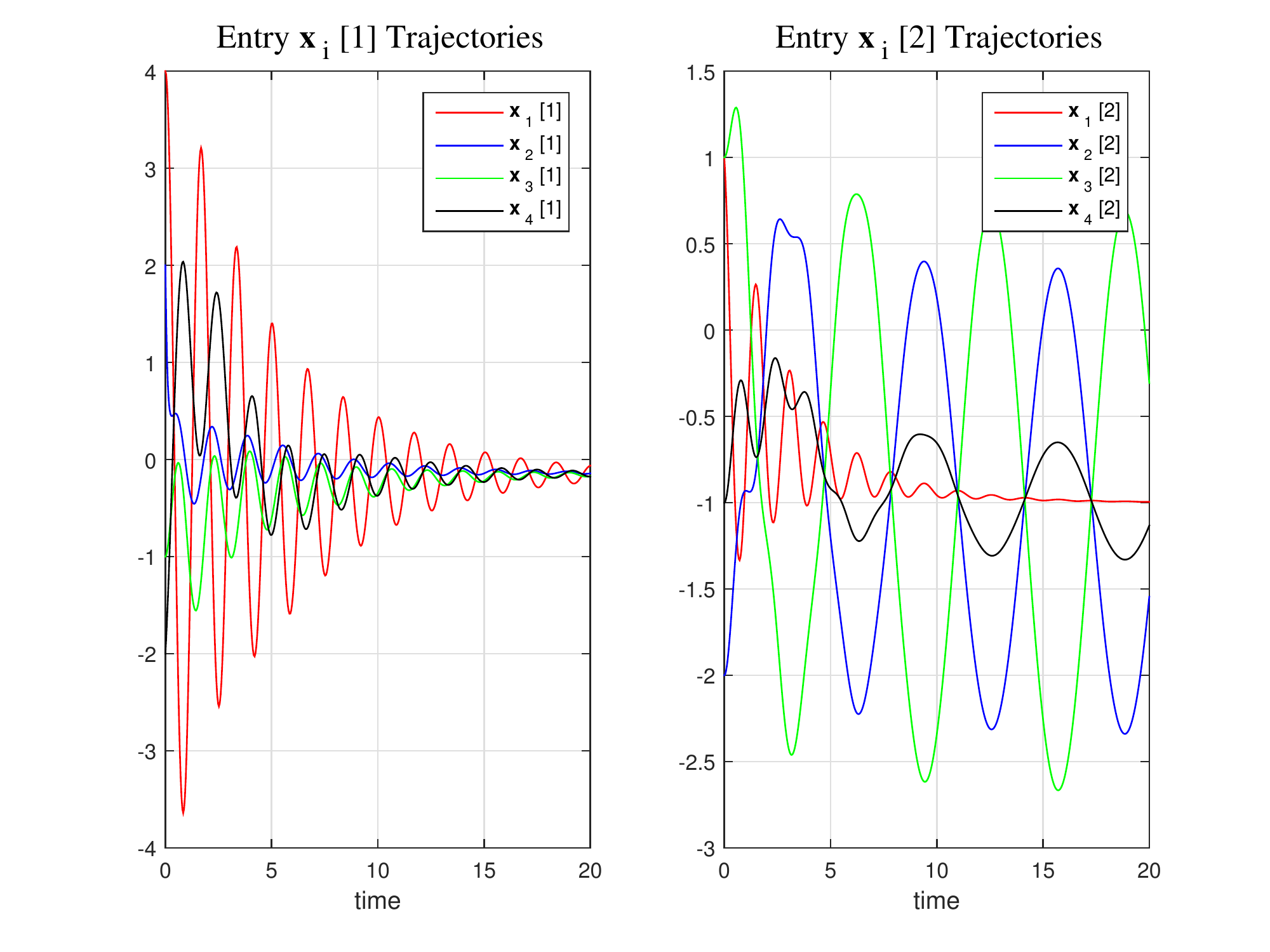}
\caption{The trajectories of $\xb_i[1]$ and $\xb_i[2]$ for $i=1,2,3,4$ over the graph in Figure \ref{fig:div_graph}. We can see that for all $i$ the first components $\xb_i[1]$ converge to $-0.1429$, while only the second component $\xb_1[2]$ converges to $-1$.}
\label{fig:div}
\end{figure}

\noindent{\bf Example 2}
Consider the linear equation ($\ref{eq:linear_equation}$) with the same $\Hb$ and $\zb$ as in Example 1. Let the interaction graph $\mathcal{G}=(\mathcal{V},\mathcal{E})$, which is a star graph studied in Section 3.3, be given in Figure \ref{fig:div_graph}. Let $\xb(0)=[4 \ 1 \ 2 \ -2 \ -1 \ 1 \ -2 \ -1]^\top$, $\vb(0)=0$ be the initial condition. As plotted in Figure \ref{fig:div}, the first entries $\xb_i[1],i=1,2,3,4$ converge to $\yb^\ast[1]=-0.1429$. However, among the second entries only $\xb_1[2]$ converges to $\yb^\ast[2]=-1$, while $\xb_2[2]$, $\xb_3[2]$ and $\xb_4[2]$ do not converge. In fact, the eigenvectors of the Laplacian for the given graph $\mathcal{G}$ are $[-0.5 \ -0.5 \ -0.5 \ -0.5]^\top$, $[0 \ -0.64 \ -0.11 \ 0.75]^\top$, $[0 \ -0.5 \ 0.8 \ -0.3]^\top$ and $[-0.87 \ 0.29 \ 0.29 \ 0.29]^\top$ corresponding to eigenvalues $0$, $1$, $1$ and $4$. For the eigenvector $\alphab=[0 \ -0.64 \ -0.11 \ 0.75]^\top$ (or $\alphab=[0 \ -0.5 \ 0.8 \ -0.3]^\top$), there holds $\dim\{\hb_i:\alphab[i]\ne 0\}=\dim\{\hb_2,\hb_3,\hb_4\}=1$, leading to oscillating trajectories for $\xb_2[2]$, $\xb_3[2]$ and $\xb_4[2]$.

\section{Discrete-time Algorithm} \label{sec:4}
In this section, we investigate the discrete-time analog of the flow (\ref{eq:flowModel1}). We index time as $k=0,1,2,\dots$ and propose the following algorithm:
\begin{equation}
\begin{aligned}
\xb(k+1) &=  \xb(k)- \epsilon(\Lb\otimes \Ib_m) \vb(k) - \epsilon\nabla U(\xb(k)) \\
\vb(k+1) &= \vb(k) + \epsilon(\Lb\otimes \Ib_m) \xb(k). \label{eq:flowModel_discrete1}
\end{aligned}
\end{equation}
For $[\xb_i^\top(k) \ \vb_i^\top(k)]^\top$ held by node $i$, (\ref{eq:flowModel_discrete1}) gives
\begin{equation}
\begin{aligned}
\xb_i(k+1) &=  \xb_i(k)- \epsilon\sum_{j\in\mathcal{N}_i}(\vb_i(k)-\vb_j(k))\\
&- \epsilon(\hb_i\hb_i^\top\xb_i(k)-z_i\hb_i)\\ \notag
\vb_i(k+1) &= \vb_i(k) + \epsilon\sum_{j\in\mathcal{N}_i}(\xb_i(k)-\vb_i(k)). \notag
\end{aligned}
\end{equation}
Therefore, the algorithm (\ref{eq:flowModel_discrete1}) inherits the same distributed structure as the flow (\ref{eq:flowModel1}). Note that (\ref{eq:flowModel_discrete1}) is an Euler approximation of (\ref{eq:flowModel1}). However, since dynamical system (\ref{eq:flowModel1}) does not have all its modes exponentially stable, we cannot immediately conclude that for a sufficiently small $\epsilon$, the solution to (\ref{eq:flowModel_discrete1}) will converge to the same consensus as (\ref{eq:flowModel1}).

\subsection{Convergence Result}
Recall that $\yb^\ast$ is the unique least squares solution of (\ref{eq:linear_equation}) and denote
\[
\Mb =
\begin{bmatrix}
-\tilde{\Hb} & -(\Lb\otimes\Ib_m) \\
\Lb\otimes\Ib_m & 0
\end{bmatrix}.
\]
The following result holds.
\begin{thm}\label{thm:2}
Suppose $\spa\{\hb_i:\alphab[i]\ne 0\}=\R^m$ for all the eigenvectors $\alphab=\big[\alphab[1] \ \dots \ \alphab[N]\big]^\top\in\C^N$ of $\Lb$. Then there exists a positive constant $\epsilon^\ast$ such that
\begin{enumerate}[(i)]
\item If $0<\epsilon<\epsilon^\ast$, then along (\ref{eq:flowModel_discrete1}) we have
\[
\lim_{k\to\infty}\xb_i(k)=y^\ast, i=1,\dots,N
\]
which converge exponentially for all $i$. In this case $\vb(k)$ continues to converge to a constant.
\item If $\epsilon>\epsilon^\ast$, then along (\ref{eq:flowModel_discrete1}) there exist initial values $\xb(0)$ and $\vb(0)$ under which $[\xb(k)^\top \ \vb(k)^\top]^\top$ diverges.
\end{enumerate}
Define $\sigma^\ast(\Mb)\subset\sigma(\Mb)$ by $\sigma^\ast(\Mb):=\{\lambda\in\sigma(\Mb):\Re(\lambda)\ne 0\}$. Then $\epsilon^\ast=\min\limits_{\lambda\in\sigma^\ast(\Mb)}\big[-\frac{2\Re(\lambda)}{|\lambda|^2}\big]$.
\end{thm}
\noindent{\em Proof.}
Let $\xb^\ast=\mathbf{1}\otimes\yb^\ast$ and $\vb^\ast$ satisfy $\nabla U(\xb^\ast)+(\Lb\otimes\Ib_m)\vb^\ast=0$. We continue to use the change of variables defined by $\hat{\xb}(k)=\xb(k)-\xb^\ast$ and $\hat{\vb}(k)=\vb(k)-\vb^\ast$ so that the equilibrium of (\ref{eq:flowModel_discrete1}) is shifted. We have
\begin{align}
\hat{\xb}(k+1) &= \hat{\xb}(k) - \epsilon(\Lb\otimes\Lb_m)\hat{\vb}(k)-\epsilon\tilde{\Hb}\hat{x}(k) \notag \\
\hat{\vb}(k+1) &= \hat{\vb}(k) + \epsilon(\Lb\otimes\Ib_m)\hat{\xb}(k), \notag
\end{align}
or equivalently,
\[
\begin{bmatrix}
\hat{\xb}(k+1) \\
\hat{\vb}(k+1)
\end{bmatrix}
=
(\Ib+\epsilon\Mb)
\begin{bmatrix}
\hat{\xb}(k) \\
\hat{\vb}(k)
\end{bmatrix}.
\]
Now note that the proof of Theorem \ref{thm:1} in effect proves certain properties of the continuous time  equation
\[
\begin{bmatrix}
\dot{\hat{\xb}} \\
\dot{\hat{\vb}}
\end{bmatrix}
=
\Mb
\begin{bmatrix}
\hat{\xb} \\
\hat{\vb}
\end{bmatrix}.
\]
In fact, from the convergence properties of $[\hat{\xb}(t)^\top \ \hat{\vb}(t)^\top]^\top$ that are established in the proof of Theorem \ref{thm:1}, we can conclude the following for $\Mb$ from basic linear system theory when $\spa\{\hb_i:\alphab[i]\ne 0\}=\R^m$ for all eigenvalues $\alphab=\big[\alphab[1] \ \dots \ \alpha[N]\big]^\top$ of $\Lb$:

(a) $\Re(\lambda)<0$ for all $\lambda\in\sigma(\Mb)$ with $\lambda\ne 0$;

(b) Zero is an eigenvalue of $\Mb$ with equal algebraic and geometric multiplicity;

(c) The zero eigenspace of $\Mb$ is given by $\{[\xb^\top \ \vb^\top]^\top:(\Lb\otimes\Ib_m)\xb=0,\tilde{\Hb}\xb+(\Lb\otimes\Ib_m)\vb=0\}$.

It is straightforward that
\[
\sigma(\Ib+\epsilon\Mb) = \{1+\epsilon\lambda:\lambda\in\sigma(\Mb)\}.
\]
and then the eigenvalues of $M$, coupled with  the continuity of $\sigma(1+\epsilon\Mb)$ as a function of $\epsilon$, imply that there exists $\epsilon^\ast>0$ such that

\noindent(\lowercase\expandafter{\romannumeral 1}) when $0<\epsilon<\epsilon^\ast$, there hold
\begin{itemize}
\item $|\lambda|<1$ for all $\lambda\in\sigma(1+\epsilon\Mb)$ with $\lambda\ne 1$;
\item $1$ is an eigenvalue of $1+\epsilon\Mb$ with equal algebraic and geometric multiplicity.
\end{itemize}
Moreover, the eigenspace of $1+\epsilon\Mb$ corresponding to eigenvalue one is the same as the  eigenspace of $\Mb$ corresponding to eigenvalue zero.

Consequently, $[\hat{\xb}(k)^\top \ \hat{\vb}(k)^\top]^\top$ converges to a vector in $\R^{2Nm}$, which implies, together with the structure of the eigenspace for the eigenvalue $1$, the desired convergence for $[\xb(k)^\top \ \vb(k)^\top]^\top$.

\noindent(\lowercase\expandafter{\romannumeral 2}) when $\epsilon>\epsilon^\ast$, there exists $\lambda\in\sigma(1+\epsilon\Mb)$ with $|\lambda|>1$. Therefore, $[\hat{\xb}(k)^\top \ \hat{\vb}(k)^\top]^\top$ will diverge for certain initial values, so in turn $[\xb(k)^\top \ \vb(k)^\top]^\top$ will also diverge.

Finally, we compute the value of $\epsilon^\ast$. Consider the following set of functions of $\epsilon$: $\sigma_\lambda(\epsilon)=1+\epsilon(\Re(\lambda)+\imath\Im(\lambda))$ with $\lambda\in\sigma^\ast(\Mb)$. According to the definition of $\epsilon^\ast$, $\epsilon^\ast$ must be the smallest $\epsilon^\ast_\lambda$ for which
\[
|\sigma_\lambda(\epsilon^\ast_\lambda)|=\sqrt{(\epsilon_\lambda^\ast\Re(\lambda)+1)^2+(\epsilon_\lambda^\ast\Im(\lambda))^2}=1.
\]
Therefore, we conclude $\epsilon^\ast=\min\limits_{\lambda\in\sigma^\ast(\Mb)}\big[-\frac{2\Re(\lambda)}{|\lambda|^2}\big]$.

We have now completed the proof of Theorem \ref{thm:2}.

When $\epsilon=\epsilon^\ast$, of course $1+\epsilon^\ast\Mb$ might have complex eigenvalues on the unit circle, leading to the possibility of periodic trajectories for $[\xb(k)^\top \ \vb(k)^\top]^\top$. Based on Theorem \ref{thm:1}. (\lowercase\expandafter{\romannumeral 2}), one might also expect that periodic trajectories could occur in discrete time when the dimensionality condition is fulfilled. In contrast however, we have the following result.
\begin{thm}\label{thm:3}
If $\dim(\spa\{\hb_i:\alphab[i]\ne 0\})<m$, then for any $\epsilon>0$, there always exist trajectories $\xb(k)$ for the algorithm (\ref{eq:flowModel_discrete1})  that diverge as $k$ tends to infinity.
\end{thm}
\noindent{\em Proof.}
From the proof of Theorem \ref{thm:1} we see that $\Mb$ will have eigenvalues on the imaginary axis with $\dim(\spa\{\hb_i:\alphab[i]\ne 0\})<m$. Such $\Mb$ generate eigenvalues $\lambda\in\sigma(1+\epsilon\Mb)$ with $|\lambda|>1$. Therefore, for any $\epsilon>0$ and for all but a thin set of initial values, the trajectory $[\xb(k)^\top \ \vb(k)^\top]^\top$ diverges. This proves the desired result.
\hfill$\square$

\begin{figure}
\centering
\includegraphics[width=3in]{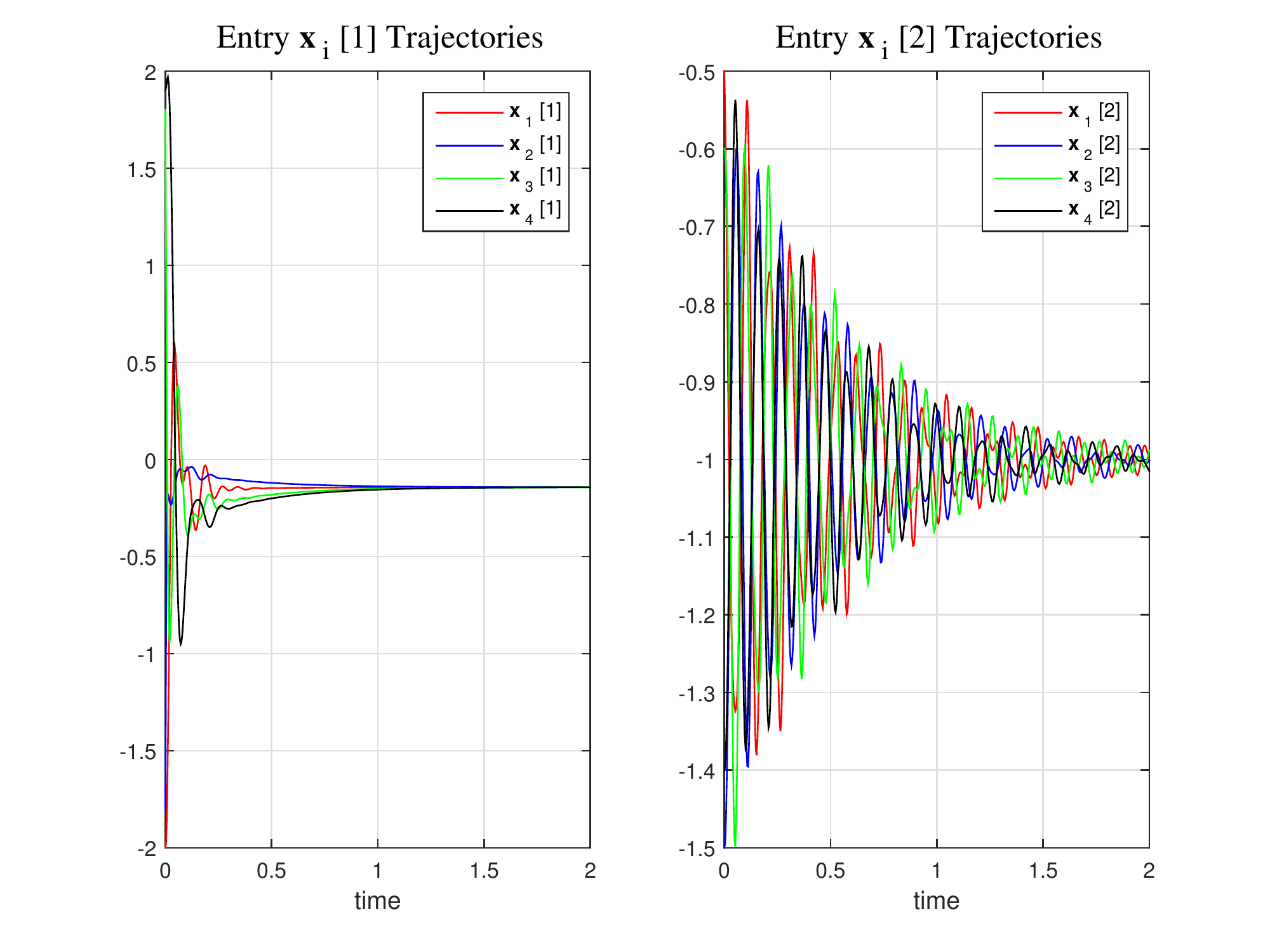}
\caption{The trajectories of $\xb_i[1]$ and $\xb_i[2]$ for $i=1,2,3,4$ over the graph in Figure \ref{fig:conv_graph} when $\epsilon=0.03$. We see that $\xb_i(t)$ converges for each $i$ to the least squares solution $[-0.1429 \ -1]^\top$.}
\label{fig:conv_small_ep}
\end{figure}

\begin{figure}
\centering
\includegraphics[width=3in]{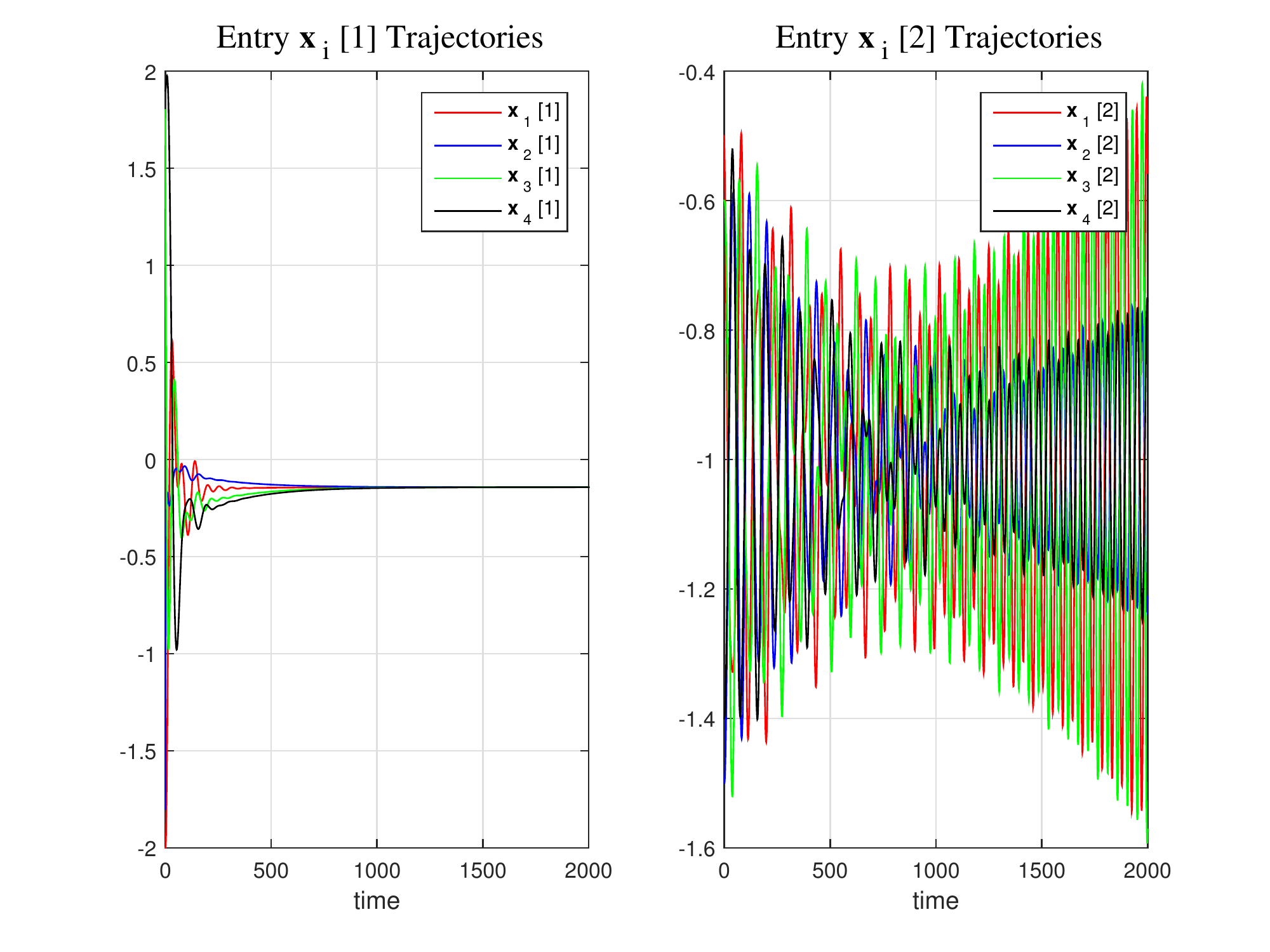}
\caption{The trajectories of $\xb_i[2]$ for $i=1,2,3,4$ diverge fast over the graph in Figure \ref{fig:conv_graph} when $\epsilon=0.04$.}
\label{fig:conv_large_ep}
\end{figure}

\subsection{Numerical Examples}
We now provide several numerical illustrations to the results stated in Theorem \ref{thm:2} and Theorem \ref{thm:3}.

\noindent {\bf Example 3.}
Consider a linear equation in the form of (\ref{eq:linear_equation}) where $\yb\in\R^2$, and
\[
\Hb =
\begin{bmatrix}
0 & 1 \\
3 & 0 \\
2 & 0 \\
1 & 0 \\
\end{bmatrix},
\quad
\zb =
\begin{bmatrix}
-1 \\ 0 \\ -2 \\ 2
\end{bmatrix}.
\]
It has a unique least squares solution $\yb^\ast=[-0.1429 \ -1]^\top$. Let $\mathcal{G}=(\mathcal{V},\mathcal{E})$ be the graph given in Figure \ref{fig:conv_graph}. By simple computation, $\epsilon^\ast=0.0362$. We have noted in Example 1 that graph $\mathcal{G}$ satisfies the condition that $\spa\{\hb_i:\alphab[i]\ne 0\}=\R^m$ for all eigenvalues $\alphab=\big[\alphab[1] \ \dots \ \alphab[N]\big]^\top$ of $\Lb$. We first set $\epsilon=0.03<\epsilon^\ast$ and plot the trajectories in Figure \ref{fig:conv_small_ep}. We can see that they all converge to the solution $\yb^\ast$. Next we set $\epsilon=0.04>\epsilon^\ast$ and plot the trajectories in Figure \ref{fig:conv_large_ep}. It is apparent that the trajectories of $\xb_i[2]$ for $i=1,2,3,4$ diverge. This example is consistent with Theorem \ref{thm:2}.

\begin{figure}
\centering
\includegraphics[width=3in]{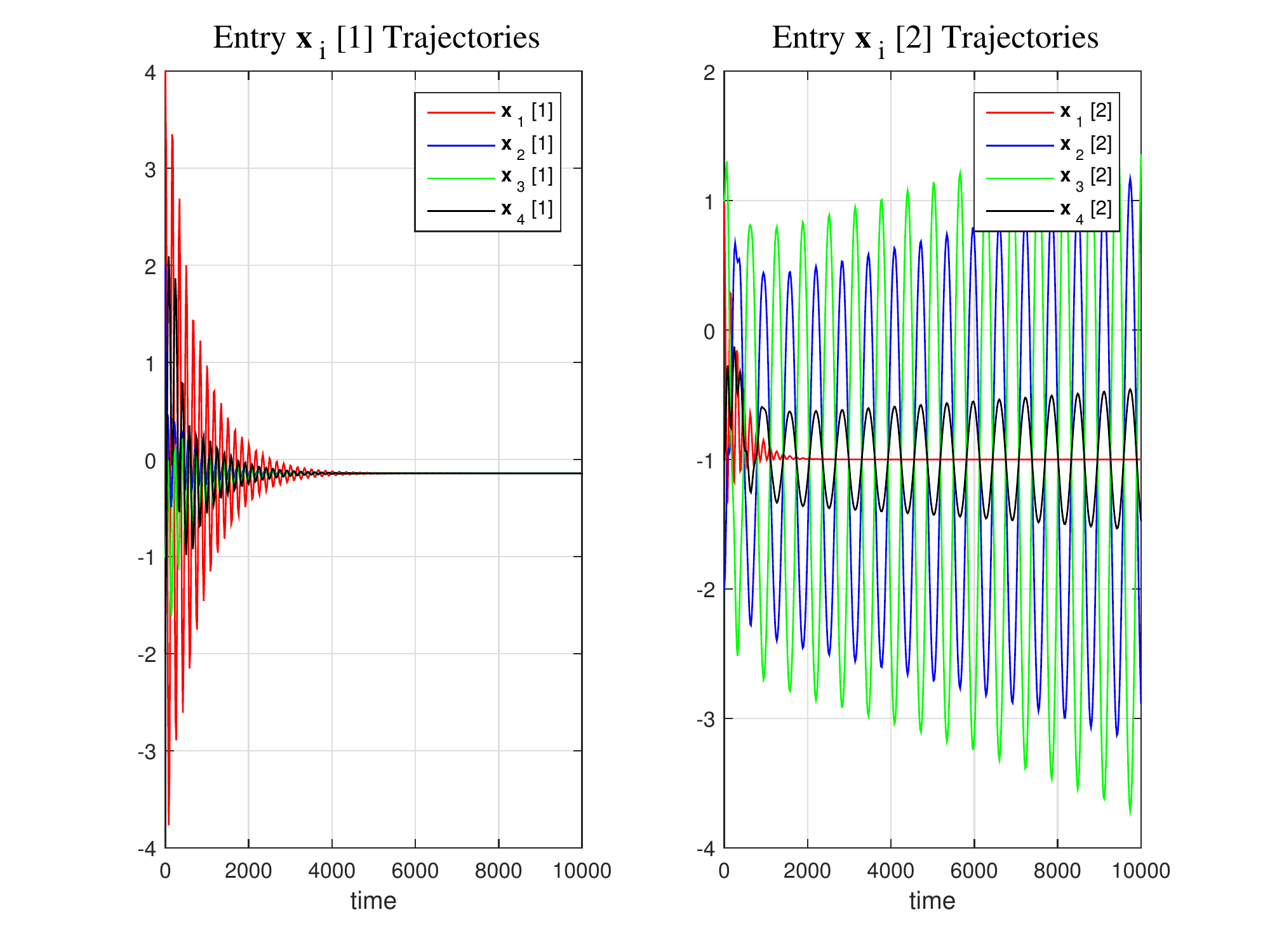}
\caption{The trajectories of $\xb_i[1]$ and $\xb_i[2]$ for $i=1,2,3,4$ over the graph in Figure \ref{fig:div_graph} when $\epsilon=0.01$. We see that the trajectories of $\xb_2[2]$, $\xb_3[2]$ and $\xb_4[2]$ diverge.}
\label{fig:div_ep}
\end{figure}

\noindent{\bf Example 4.}
 Consider the linear equation (\ref{eq:linear_equation}) with the same $\Hb$, $\zb$ and $\yb^\ast$ as in Example 3. Let $\mathcal{G}=(\mathcal{V},\mathcal{E})$ be the graph in Figure \ref{fig:div_graph} satisfying the condition of Theorem \ref{thm:3}, as noted when we explored Example 2. Let $\epsilon=0.01$. We plot the trajectories of $\xb_i[1]$ and $\xb_i[2]$ for $i=1,2,3,4$ in Figure \ref{fig:div_ep}. It is evident that the trajectories of $\xb_2[2]$, $\xb_3[2]$ and $\xb_4[2]$ diverge; this is consistent with Theorem \ref{thm:3}.

When $\epsilon=0.5$, all trajectories over both graphs in Figure \ref{fig:conv_graph} and Figure \ref{fig:div_graph} fail to converge.

\section{Further Discussions} \label{sec:5}
In this section, we provide some further results and investigations  regarding the least squares residual vector, alternative state expansion methods, and the influence of switching network structures.
\subsection{Residual vector}
The residual vector of (\ref{eq:linear_equation}) is defined as $$
\mathbf{e}^\ast=\Hb\yb^\ast-\zb
$$ with $\yb^\ast$ being the least squares solution. Since each node $i$ holds the information of $\hb_i$ and $\zb_i$, the $i$-th component $e_i^\ast$ of the residual vector $\mathbf{e}^\ast$ can be obtained by node $i$ after running the continuous-time flow (\ref{eq:flowModel1}) by $e_i^\ast=\lim\limits_{t\to\infty}(\hb_i^\top\xb_i(t)-z_i)$ if the condition of Theorem \ref{thm:1}.(\lowercase\expandafter{\romannumeral 1}) is satisfied. Therefore, after executions of the presented continuous or discrete algorithms,   another  distributed agreement algorithm, e.g., \cite{pease1980reaching}, can be applied to the network so that all nodes can learn  the entire residual vector $\mathbf{e}^\ast$.

\subsection{State Expansion}
Consider a linear equation with respect to $\bar{\yb}$:
\begin{equation}\label{eq:higherDim}
\bar{\Hb}\bar{\yb}=\bar{\zb}
\end{equation}
where
\[
\bar{\Hb} =
\begin{bmatrix}
\Hb & -\Ib_N \\
0 & \Hb^\top
\end{bmatrix}
\]
and $\bar{\zb} = [\zb^\top \ 0]^\top$. It is known and easily checked that if $\Hb$ has full column rank, (\ref{eq:higherDim}) has a unique exact solution in the form of $\bar{\yb}^\ast = [\yb^{\ast\top} \ \mathbf{e}^{\ast\top}]^\top$ where $\yb^\ast$ is the least squares solution of (\ref{eq:linear_equation}) and $\mathbf{e}^\ast$ is the residual vector. Therefore, distributed algorithms presented in \cite{mou2013fixed,mou2015distributed,liu2013asynchronous,lu2009distributed1,lu2009distributed2,anderson2015decentralized,shi:15} for solving linear algebraic equations with exact solutions can be used to solve (\ref{eq:higherDim}), which in turn gives us the least squares solution of (\ref{eq:linear_equation}).

Such algorithms admit simple first-order structure even working for time-varying networks, e.g.,  \cite{mou2015distributed,shi:15}. However, in order to construct (\ref{eq:higherDim}), we have to add some additional $m$ nodes to the network which have access to the columns of $\Hb$, imposing  a much more restricted assumption on the node information structure.


\subsection{Switching Networks}
We have known that the network flow (\ref{eq:flowModel1}) defines a linear time-invariant (LTI) system since the network is fixed. However, many networks from real-world applications are time-varying, e.g.,  \cite{neely2005dynamic,ahmed2009recovering}. If the network underlying (\ref{eq:flowModel1}) is indeed switching, (\ref{eq:flowModel1}) will in turn induce linear time-varying (LTV) dynamics. Since (\ref{eq:flowModel1}) is only marginally stable regardless of choice of network, generally speaking,  existing theories for the stability of LTV systems cannot be directly applied. In this section, we use numerical examples to study the asymptotic properties of (\ref{eq:flowModel1}) under switching networks.

\begin{figure}
\centering
\includegraphics[width=1.4in]{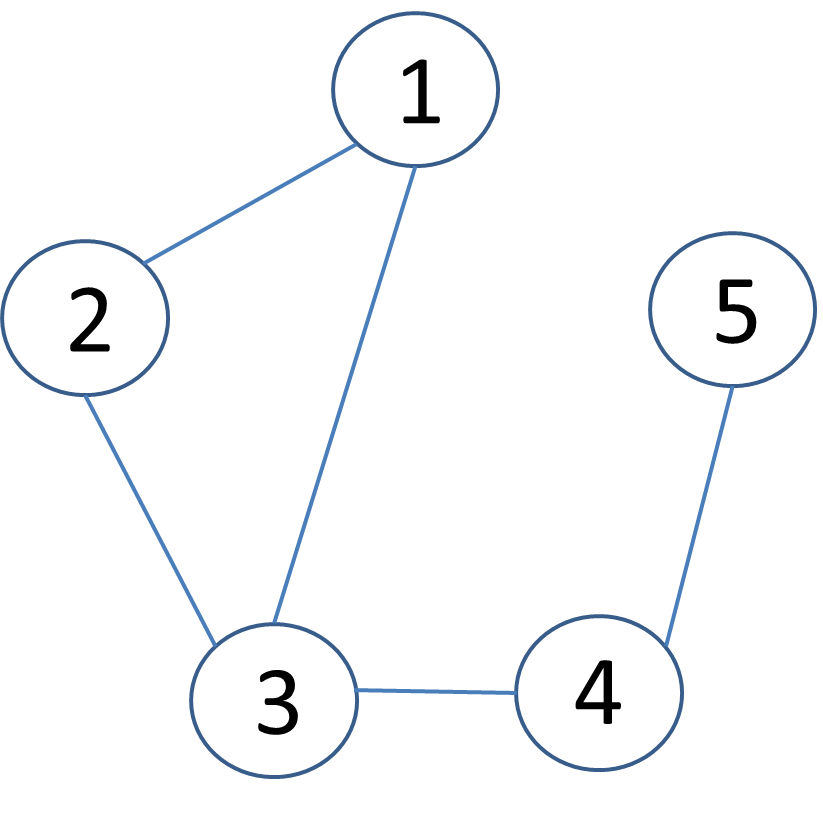}
\caption{Graph $\mathcal{G}_1$ with five nodes.}
\label{fig:l1}
\end{figure}

\begin{figure}
\centering
\includegraphics[width=1.4in]{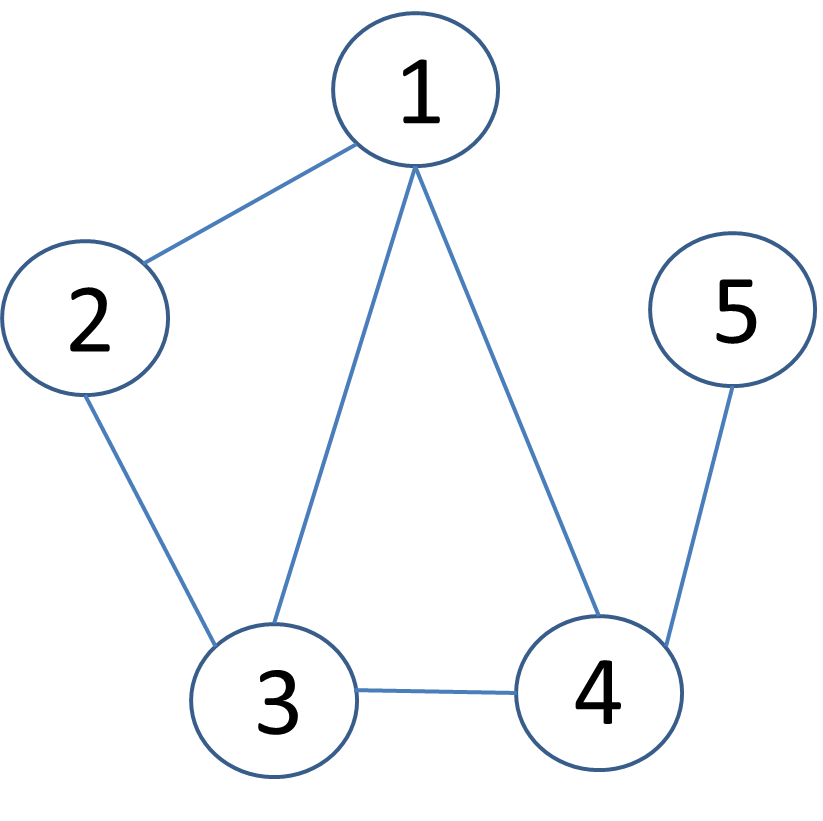}
\caption{Graph $\mathcal{G}_2$ with five nodes.}
\label{fig:l2}
\end{figure}

We now define precisely networks with switching interaction structures. Let $\mathcal{Q}$ be the set containing all graphs with node set $\mathcal{V}$. Let $\mathcal{Q}^\ast\subset\mathcal{Q}$ and introduce a piecewise constant mapping $\mathcal{G}_{\sigma(\cdot)}:\R^{\ge 0}\to\mathcal{Q}^\ast$. Then $\mathcal{G}_{\sigma(t)}=(\mathcal{V},\mathcal{E}_{\sigma(t)})$ represents the network topology at time $t$. We consider a network with $5$ nodes indexed in $\mathcal{V}=\{1,2,3,4,5\}$. We also let $\mathcal{Q}^\ast=\{\mathcal{G}_1,\mathcal{G}_2\}$ where $\mathcal{G}_1$, $\mathcal{G}_2$ are as shown in Fig. \ref{fig:l1}, Fig. \ref{fig:l2}, respectively. Denote the Laplacian of $\mathcal{G}_1$, $\mathcal{G}_2$ by $\Lb_1$, $\Lb_2$. We know from direct calculation that $\mathcal{I}_{\alphab}=\{1,2,3,4,5\}$ or $\mathcal{I}_{\alphab}=\{1,2\}$ for $\alphab\in\mathcal{S}_{\Lb_1}$ and $\mathcal{I}_{\alphab}=\{1,2,3,4,5\}$ or $\mathcal{I}_{\alphab}=\{1,3\}$ for $\alphab\in\mathcal{S}_{\Lb_2}$. For simplicity we assume that the switchings in the graph signal are periodic, as indicated in the following:
$$\mathcal{G}_{\sigma(t)}=\left\{
\begin{aligned}
& \mathcal{G}_1,\ t\in\big [T l,T (l+1)\big ), l=0,2,4,\dots \\
& \mathcal{G}_2,\ t\in\big [T l,T (l+1)\big ), l=1,3,5,\dots \\
\end{aligned}
\right.
$$
where $T\in\R$ is the period.

\subsubsection{Switching of Converging Networks}
First of all, we study the case where the graphs in $\mathcal{Q}^\ast$ satisfy the condition stated in Theorem \ref{thm:1}.(\lowercase\expandafter{\romannumeral 1}), i.e.,  each individual graph in the switching graph signals will be able to produce trajectories for all $\mathbf{x}_i(t)$ converging to the least square solutions in the absence of switching.

\begin{figure*} [htbp]
\subfigure[T=100]{
\begin{minipage}{0.33\linewidth}
\centering
\includegraphics[width=5.2cm]{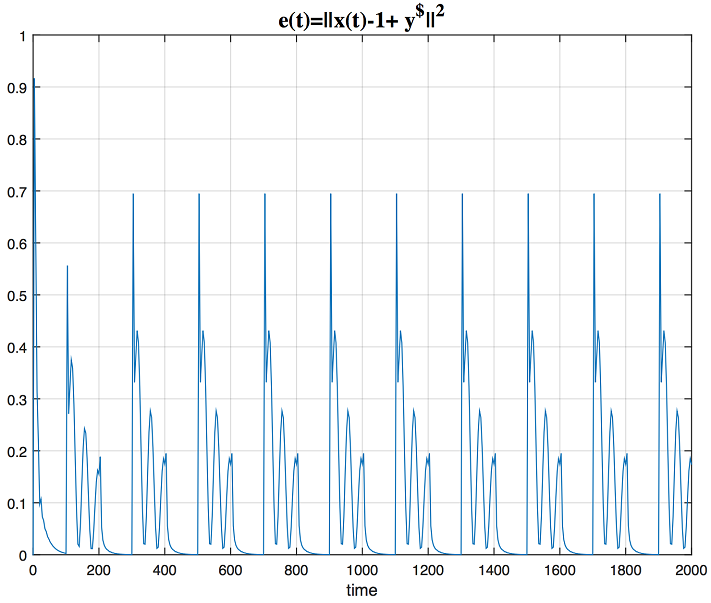}
\end{minipage}
}
\subfigure[T=10]{
\begin{minipage}{0.33\linewidth}
\centering
\includegraphics[width=5.2cm]{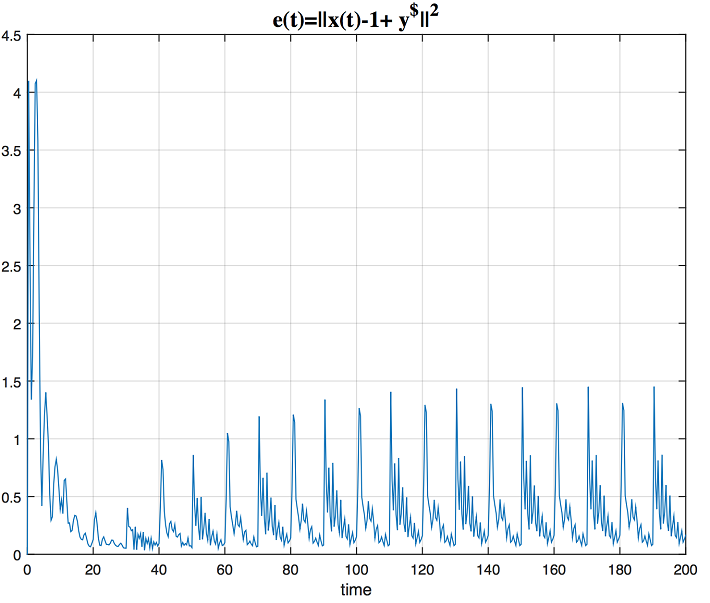}
\end{minipage}
}
\subfigure[T=1]{
\begin{minipage}{0.33\linewidth}
\centering
\includegraphics[width=5.2cm]{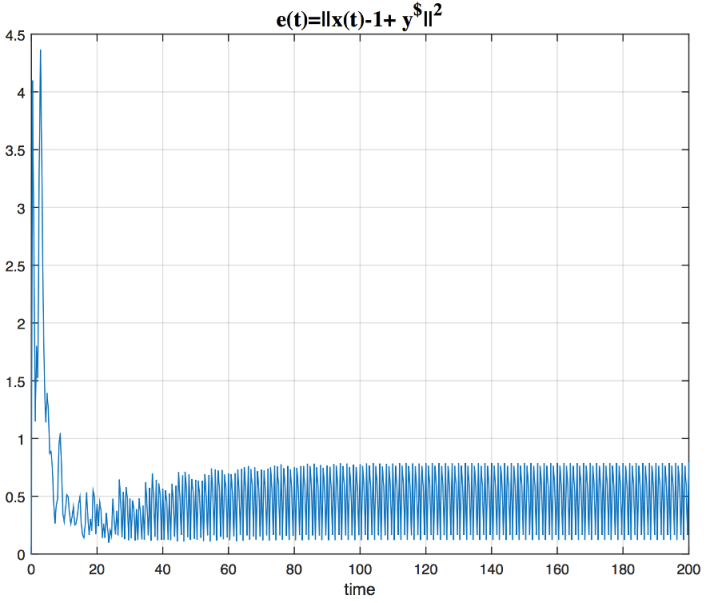}
\end{minipage}
}
\caption{The trajectories of $e(t)=\|\xb(t)-\mathbf{1}_5\otimes\yb^\ast\|^2$ for $i=1,2,3,4,5$ when (a) $T=100$, (b) $T=10$, (c) $T=1$.}
\label{fig:conv600_s5m2_d}
\end{figure*}

\medskip

\begin{figure*}
\subfigure[T=100]{
\begin{minipage}{0.33\linewidth}
\centering
\includegraphics[width=5.2cm]{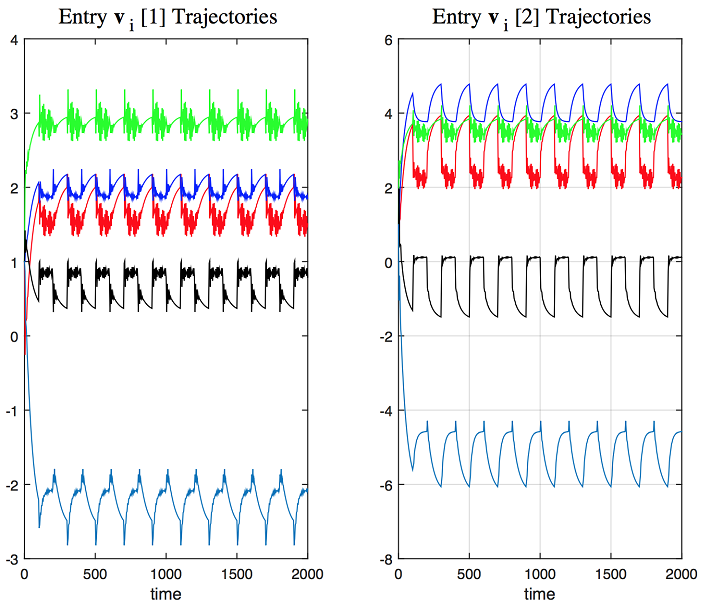}
\end{minipage}
}
\subfigure[T=10]{
\begin{minipage}{0.33\linewidth}
\centering
\includegraphics[width=5.2cm]{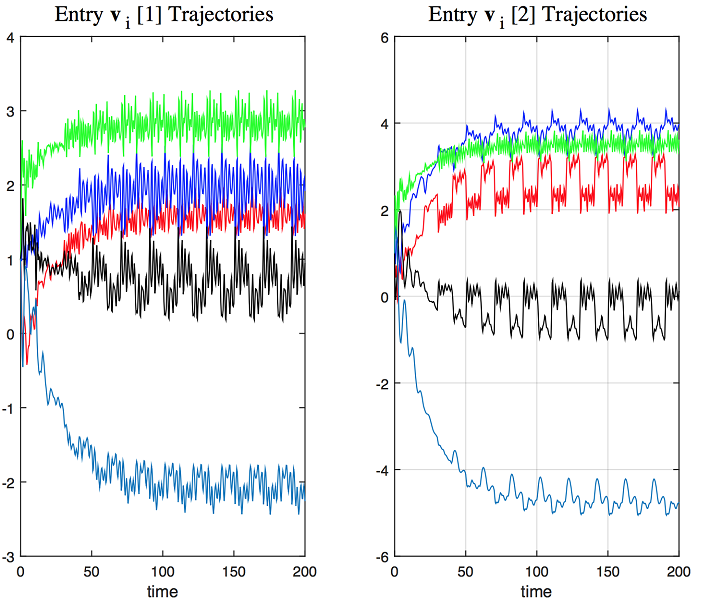}
\end{minipage}
}
\subfigure[T=1]{
\begin{minipage}{0.33\linewidth}
\centering
\includegraphics[width=5.2cm]{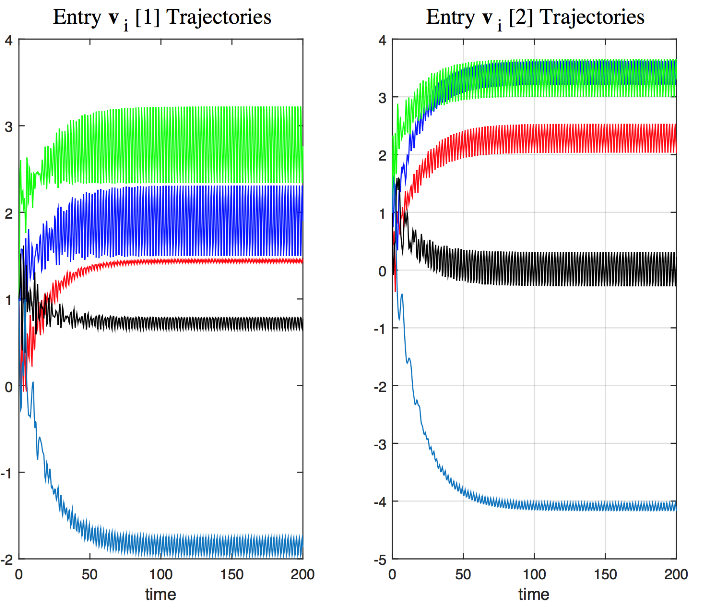}
\end{minipage}
}
\caption{The trajectories of $(\vb_i(t))[1]$ and $(\vb_i(t))[2]$ for $i=1,2,3,4,5$ when (a) $T=100$, (b) $T=10$, (c) $T=1$.}
\label{fig:conv600_s5m2_v}
\end{figure*}


\noindent {\bf Example 5.} Consider a linear equation (\ref{eq:linear_equation}) with
\[
\Hb =
\begin{bmatrix}
3 & 2 \\
1 & -3 \\
1 & 1 \\
-1.5 & 4 \\
2.5 & 4
\end{bmatrix}, \
\zb =
\begin{bmatrix}
2 \\
1 \\
5 \\
-2.5 \\
0.25
\end{bmatrix}.
\]
Then it has a unique least squares solution as $\yb^\ast=[0.93 \ -0.23]^\top$. It can be verified that the condition of Theorem \ref{thm:1}.(\lowercase\expandafter{\romannumeral 1}) is met for both $\mathcal{G}_1$ and $\mathcal{G}_2$. Let $\xb(0) = [1\ -0.5 \ 1.3 \ -0.8 \ 0.7 \ 0.6 \ 0.7\ -1.4 \ -0.5 \ 1]^\top$, $\vb(0) = \mathbf{1}_{10}$ be the initial condition. In Figure \ref{fig:conv600_s5m2_d}, we plot the trajectories of $$
e(t)=\|\xb(t)-\mathbf{1}_5\otimes\yb^\ast\|^2
$$ for $T=100,10,1$, respectively. We find that $e(t)$ oscillates at the same frequency as $\mathcal{G}_{\sigma(t)}$ for all these three cases. Then the trajectories of $(\vb_i(t))[1]$, $(\vb_i(t))[2]$, $i=1,2,3,4,5$ for $T=100,10,1$, respectively, are plotted in Figure \ref{fig:conv600_s5m2_v}. It turns out that $\vb(t)$ admits similar oscillating trajectories.


Let $\vb_1^\ast$, $\vb_2^\ast$ be the solutions corresponding to $\vb^\ast$ of (\ref{eq:equilib}) with Laplacian $\Lb$ defined over the graphs $\mathcal{G}_1$, $\mathcal{G}_2$, respectively. Let $\mathbf{W}_{\mathcal{G}_1}$, $\mathbf{W}_{\mathcal{G}_2}$ be defined as in the proof of Theorem \ref{thm:1}. Define $\mathcal{K}_1$, $\mathcal{K}_2$ as
\begin{equation}\notag
\begin{aligned}
\mathcal{K}_1 &= \{(\Ib_{10}-\mathbf{W}_{\mathcal{G}_1})\vb^\ast_1 + \mathbf{W}_{\mathcal{G}_1}\kb:\kb\in\R^{10}\}, \\
\mathcal{K}_2 &= \{(\Ib_{10}-\mathbf{W}_{\mathcal{G}_2})\vb^\ast_2 + \mathbf{W}_{\mathcal{G}_2}\kb:\kb\in\R^{10}\}.
\end{aligned}
\end{equation}
Then $\mathcal{K}_1$, $\mathcal{K}_2$ are the limit sets of $\vb(t)$ as shown in Theorem \ref{thm:1} for networks with fixed interaction graphs $\mathcal{G}_1$ and $\mathcal{G}_2$, respectively. Simple calculation can show that $\mathcal{K}_1\cap\mathcal{K}_2=\emptyset$.
Therefore, when the network switches from $\mathcal{G}_1$ to $\mathcal{G}_2$ (or, from $\mathcal{G}_2$ to $\mathcal{G}_1$), $\vb(t)$ has to transit from approaching to $\mathcal{K}_1$ ($\mathcal{K}_2$) to approaching $\mathcal{K}_2$ ($\mathcal{K}_1$). This leads to an oscillation in $\vb(t)$, which in turn causes $\xb(t)$ to oscillate.


In the seminal work of \cite{morse1996supervisory}, it was shown that if a linear switching system takes slow switching and each plant of the switching system is asymptotically stable, then the switching system will continue to be  asymptotically stable. However,  based on the above limit set arguments as well as the numerical verifications, in general  the system (\ref{eq:flowModel1}) for switching networks can only exhibit oscillatory behavior even under low switching frequencies, although each individual network yields converging trajectories of $\xb(t)$ to a common limit. This observation points to an interesting question analogous to the work of \cite{morse1996supervisory} regarding behaviors of linear time-varying systems when the plants are only marginally stable.

\subsubsection{Switching of Non-converging Networks}
Next, we study the case where all graphs in $\mathcal{Q}^\ast$ satisfy the condition of Theorem \ref{thm:1}.(\lowercase\expandafter{\romannumeral 2}), i.e., neither of  the graphs in the switching graph signal will be able to provide convergent solutions under (\ref{eq:flowModel1}).

\begin{figure*} [htbp]
\subfigure[T=0.5]{
\begin{minipage}{0.33\linewidth}
\centering
\includegraphics[width=5.2cm]{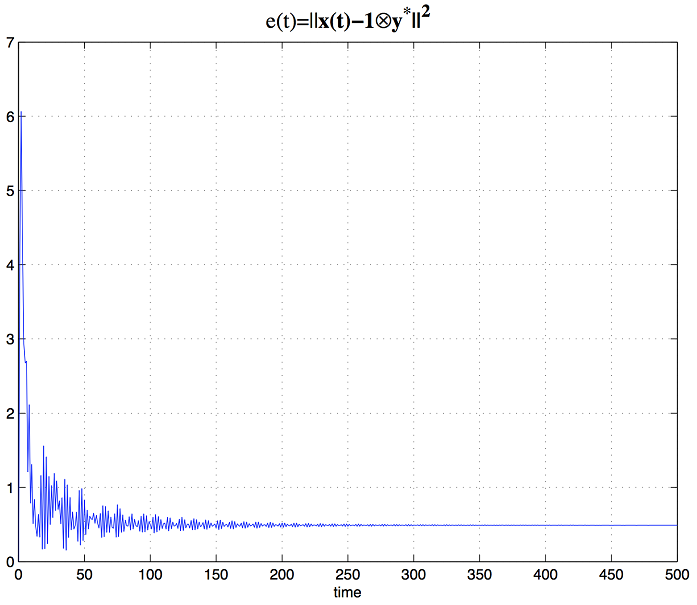}
\end{minipage}
}
\subfigure[T=0.25]{
\begin{minipage}{0.33\linewidth}
\centering
\includegraphics[width=5.2cm]{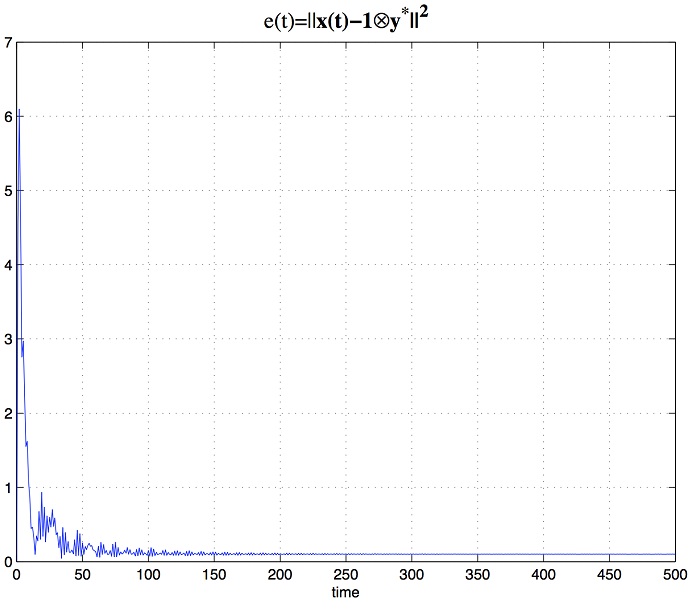}
\end{minipage}
}
\subfigure[T=0.1]{
\begin{minipage}{0.33\linewidth}
\centering
\includegraphics[width=5.2cm]{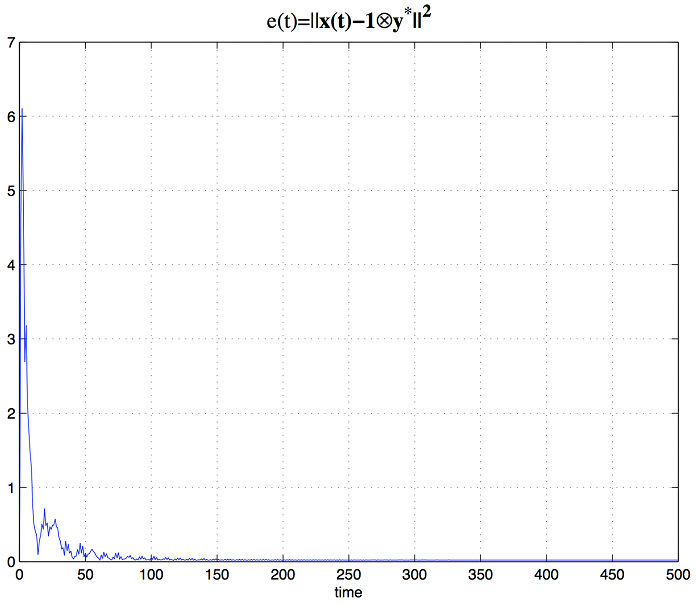}
\end{minipage}
}
\caption{The trajectories of $e(t)=\|\xb(t)-\mathbf{1}_5\otimes\yb^\ast\|^2$ for $i=1,2,3,4,5$ when (a) $T=0.5$, (b) $T=0.25$, (c) $T=0.1$.}
\label{fig:conv600_s5m3_d}
\end{figure*}



\begin{figure*} [htbp]
\subfigure[T=0.5]{
\begin{minipage}{0.33\linewidth}
\centering
\includegraphics[width=5.2cm]{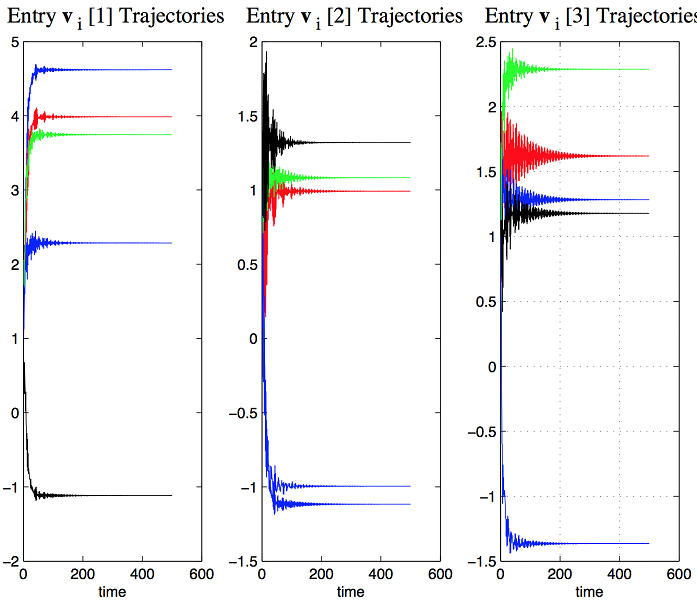}
\end{minipage}
}
\subfigure[T=0.25]{
\begin{minipage}{0.33\linewidth}
\centering
\includegraphics[width=5.2cm]{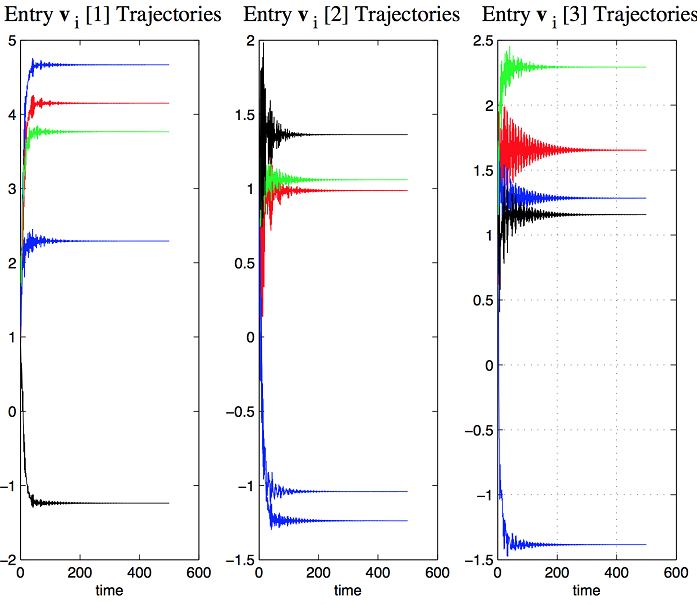}
\end{minipage}
}
\subfigure[T=0.1]{
\begin{minipage}{0.33\linewidth}
\centering
\includegraphics[width=5.2cm]{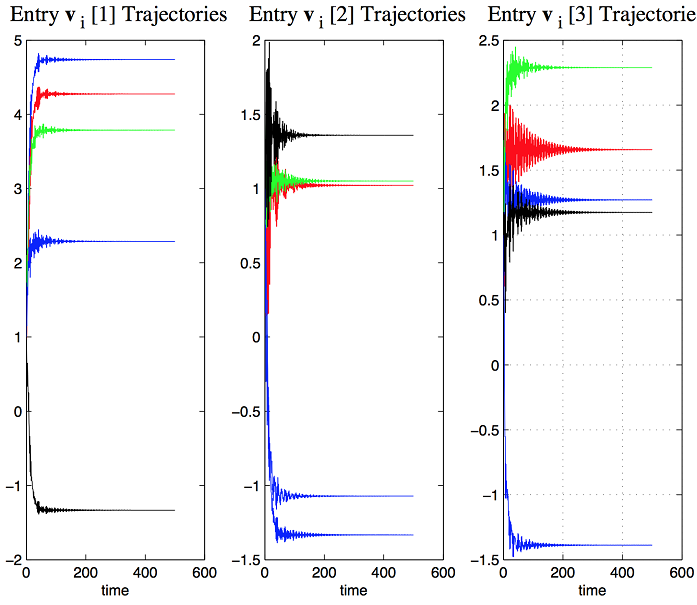}
\end{minipage}
}
\caption{The trajectories of $(\vb_i(t))[1]$, $(\vb_i(t))[2]$, $(\vb_i(t))[3]$ for $i=1,2,3,4,5$ when (a) $T=0.5$, (b) $T=0.25$, (c) $T=0.1$.}
\label{fig:conv600_s5m3_v}
\end{figure*}

\noindent {\bf Example 6.} Consider a linear equation (\ref{eq:linear_equation}) with respect to $\yb$ where
\[
\Hb =
\begin{bmatrix}
3 & 2 & 0\\
1 & -3 & -1\\
2 & 1 & 1.5\\
-7 & -2 & 3 \\
2 & -0.5 & 1
\end{bmatrix}, \
\zb =
\begin{bmatrix}
1 \\
5 \\
3 \\
-1 \\
0
\end{bmatrix}.
\]
It can be easily calculated that the unique least squares solution is $\yb^\ast=[0.75 \ -1.12 \ 0.53]^\top$. At the same time, the condition of Theorem \ref{thm:1}.(\lowercase\expandafter{\romannumeral 2}) holds for both $\mathcal{G}_1$ and $\mathcal{G}_2$.
Let $\xb(0) = [-1\ -0.5 \ 1 \ 0.8 \ -0.75 \ 0.5 \ 0.7 \ -0.6 \ -0.3 \ -0.8 \ -1.6 \ 0.25 \ 0.5 \ -1 \ 0.7]^\top$, $\vb(0) = \mathbf{1}_{15}$ be the initial condition. In Figure \ref{fig:conv600_s5m3_d} we plot the trajectories of $e(t)=\|\xb(t)-\mathbf{1}_5\otimes\yb^\ast\|^2$ for $T=0.5,0.25,0.1$, respectively. We can see  in the three cases that $e(t)$ asymptotically tends to a neighborhood of zero as $t$ goes to infinity. Furthermore,  the faster the network switches, the smaller such neighborhood becomes. Then the trajectories of $(\vb_i(t))[1]$, $(\vb_i(t))[2]$, $(\vb_i(t))[3]$, $i=1,2,3,4,5$ for $T=0.5,0.25,0.1$, respectively, are plotted in Figure \ref{fig:conv600_s5m3_v}. We find again that $\vb(t)$ tends to fall in a neighborhood of a fixed point as time grows to infinity.


Remarkably enough, neither $\mathcal{G}_1$ or $\mathcal{G}_2$ alone  will be able to produce trajectories for the $\mathbf{x}_i(t)$ that converge even to a neighborhood of the least square solutions, as proved in Theorem \ref{thm:1}. The above numerical example clearly reveals the possibility that the two graphs   $\mathcal{G}_1$ and $\mathcal{G}_2$ can {\em cooperate} by a fast switching signal, under which  the resulting node dynamics can be driven to a small neighborhood around the least square solution of Eq. (\ref{eq:linear_equation}). It will be interesting to look at the mechanism behind such a surprising phenomenon.

\vspace{5mm}

\section{Conclusions} \label{sec:6}
We studied the problem of obtaining the least squares solution to a linear algebraic equation using distributed algorithms. Each node has the information of one scalar linear equation and holds a dynamic state. Two distributed algorithms in continuous time and discrete time respectively were developed as least squares solvers for linear equations. Under certain conditions, all node states can reach a consensus, which gives the least square solution, by exchanging information with neighbors over a network. To verify the convergence result, some numerical examples were provided. Besides, the feasibility of several fundamental graphs was discussed.   Finally, methods of computing residual vectors at individual nodes and alternative state expansion method were discussed, and we also  investigated the performance of the proposed flow under switching networks. Remarkably,   switching networks at high switching frequencies define an approximate least squares solver even if all graphs in the switching signal fail to do so in the absence of structure switching. Future directions currently being contemplated include establishing the convergence rate of the distributed algorithms,  distributed identification of the residual vector, which can be of practical interest
and modifying the underlying cost function or adding constraints on it to reflect objectives such as outlier suppression.


\begin{thebibliography}{10}

\bibitem{ablowitz2003complex}
M.~J Ablowitz and A.~S. Fokas.
\newblock {\em Complex variables: introduction and applications}.
\newblock Cambridge University Press, 2003.

\bibitem{ahmed2009recovering}
A. Ahmed and E.~P. Xing.
\newblock Recovering time-varying networks of dependencies in social and
  biological studies.
\newblock {\em Proceedings of the National Academy of Sciences},
  106(29):11878--11883, 2009.

\bibitem{anderson2015decentralized}
B. Anderson, S. Mou, A.~S. Morse, and U. Helmke.
\newblock Decentralized gradient algorithm for solution of a linear equation.
\newblock {\em Numerical Algebra, Control and Optimization\emph{, to
  appear(preprint arXiv:1509.04538)}}.

\bibitem{arrow:58}
K.~J. Arrow, L. Hurwicz, and H. Uzawa.
\newblock {\em Studies in linear and non-linear programming}.
\newblock With contributions by H. B. Chenery, S. M. Johnson, S. Karlin, T.
  Marschak, R. M. Solow. Stanford Mathematical Studies in the Social Sciences,
  vol. II. Stanford University Press, Stanford, Calif., 1958.

\bibitem{ayari2016multi}
R. Ayari, I. Hafnaoui, A. Aguiar, P. Gilbert, M.
  Galibois, J.-P. Rousseau, G. Beltrame, and G. Nicolescu.
\newblock Multi-objective mapping of full-mission simulators on heterogeneous
  distributed multi-processor systems.
\newblock {\em The Journal of Defense Modeling and Simulation: Applications,
  Methodology, Technology}, page 1548512916657907, 2016.

\bibitem{bertsekas1999nonlinear}
D.~P Bertsekas.
\newblock {\em Nonlinear programming}.
\newblock Athena scientific Belmont, 1999.

\bibitem{boyd2004convex}
S. Boyd and L. Vandenberghe.
\newblock In {\em Convex optimization}, pages 215--227. Cambridge university
  press, 2004.

\bibitem{de2007distributed}
C. A. F. De~Rose, H.-U. Heiss, and B. Linnert.
\newblock Distributed dynamic processor allocation for multicomputers.
\newblock {\em Parallel Computing}, 33(3):145--158, 2007.

\bibitem{elbirt2005instruction}
A.~J. Elbirt and C. Paar.
\newblock An instruction-level distributed processor for symmetric-key
  cryptography.
\newblock {\em IEEE Transactions on Parallel and distributed Systems},
  16(5):468--480, 2005.

\bibitem{fuhrmann2015mathematics}
P.~A. Fuhrmann and U. Helmke.
\newblock {\em The mathematics of networks of linear systems}.
\newblock Springer, 2015.

\bibitem{garland2008parallel}
M. Garland, S. Le~Grand, J. Nickolls, J. Anderson, J. Hardwick,
  Scott Morton, Everett Phillips, Yao Zhang, and Vasily Volkov.
\newblock Parallel computing experiences with cuda.
\newblock {\em Micro, IEEE}, 28(4):13--27, 2008.

\bibitem{cortez:14}
B.~Gharesifard and J.~Cort{\'e}s.
\newblock Distributed continuous-time convex optimization on weight-balanced
  digraphs.
\newblock {\em IEEE Transactions on Automatic Control}, 59(3):781--786, March
  2014.

\bibitem{gharesifard2014distributed}
B. Gharesifard and J. Cort{\'e}s.
\newblock Distributed continuous-time convex optimization on weight-balanced
  digraphs.
\newblock {\em IEEE Transactions on Automatic Control}, 59(3):781--786, 2014.

\bibitem{gower2015randomized}
R.~M. Gower and P. Richt{\'a}rik.
\newblock Randomized iterative methods for linear systems.
\newblock {\em SIAM Journal on Matrix Analysis and Applications},
  36(4):1660--1690, 2015.

\bibitem{jakovetic2014fast}
D. Jakoveti{\'c}, J. Xavier, and J.~M. .................F.  Moura.
\newblock Fast distributed gradient methods.
\newblock {\em IEEE Transactions on Automatic Control}, 59(5):1131--1146, 2014.

\bibitem{kaczmarz1937angenaherte}
S. Kaczmarz.
\newblock Angen{\"a}herte aufl{\"o}sung von systemen linearer gleichungen.
\newblock {\em Bulletin International de l’Academie Polonaise des Sciences et
  des Lettres}, 35:355--357, 1937.

\bibitem{keckler2011gpus}
S.~W. Keckler, W.~J. Dally, B. Khailany, M. Garland, and D.
  Glasco.
\newblock Gpus and the future of parallel computing.
\newblock {\em IEEE Micro}, 31(5):7--17, 2011.

\bibitem{completegraph2009}
J. Kelner.
\newblock {\em Lecture 2}, 2009 (accessed 11/10/2016).
\newblock \url{https://ocw.mit.edu/courses/mathematics/}.

\bibitem{alan:kroneckerproduct}
Alan~J Laub.
\newblock In {\em Matrix analysis for scientists and engineers}, pages
  139--148. Siam, 2005.

\bibitem{liu2013asynchronous}
J.~Liu, S. Mou, and A.~S. Morse.
\newblock An asynchronous distributed algorithm for solving a linear algebraic
  equation.
\newblock In {\em 52nd IEEE Conference on Decision and Control}, pages
  5409--5414, 2013.

\bibitem{lu2009distributed1}
J. Lu and C.~Y. Tang.
\newblock Distributed asynchronous algorithms for solving positive definite
  linear equations over networks—{P}art I: Agent networks.
\newblock {\em IFAC Proceedings Volumes}, 42(20):252--257, 2009.

\bibitem{lu2009distributed2}
J. Lu and C.~Y. Tang.
\newblock Distributed asynchronous algorithms for solving positive definite
  linear equations over networks—{P}art II: Wireless networks.
\newblock {\em IFAC Proceedings Volumes}, 42(20):258--263, 2009.

\bibitem{margaris2014parallel}
A. Margaris, S. Souravlas, and M. Roumeliotis.
\newblock Parallel implementations of the jacobi linear algebraic systems
  solve.
\newblock {\em preprint arXiv:1403.5805}, 2014.

\bibitem{morse1996supervisory}
A.p~S.  Morse.
\newblock Supervisory control of families of linear set-point controllers
  {P}art I. exact matching.
\newblock {\em IEEE Transactions on Automatic Control}, 41(10):1413--1431,
  1996.

\bibitem{mou2013fixed}
S. Mou and A. S.~Morse.
\newblock A fixed-neighbor, distributed algorithm for solving a linear
  algebraic equation.
\newblock In {\em Proc. European Control Conference}, pages 2269--2273, 2013.

\bibitem{mou2015distributed}
S. Mou, J.~Liu, and A.~S. Morse.
\newblock A distributed algorithm for solving a linear algebraic equation.
\newblock {\em IEEE Transactions on Automatic Control}, 60(11):2863--2878,
  2015.

\bibitem{nedic2009distributed1}
A. Nedic and A. Ozdaglar.
\newblock Distributed subgradient methods for multi-agent optimization.
\newblock {\em IEEE Transactions on Automatic Control}, 54(1):48--61, 2009.

\bibitem{nedic2010constrained1}
A. Nedic, A. Ozdaglar, and P.~A. Parrilo.
\newblock Constrained consensus and optimization in multi-agent networks.
\newblock {\em IEEE Transactions on Automatic Control}, 55(4):922--938, 2010.

\bibitem{neely2005dynamic}
Mi.~J Neely, E.Modiano, and C.~E Rohrs.
\newblock Dynamic power allocation and routing for time-varying wireless
  networks.
\newblock {\em IEEE Journal on Selected Areas in Communications},
  23(1):89--103, 2005.

\bibitem{partl2011enabling}
A.~M. Partl, A. Maselli, B. Ciardi, A. Ferrara, and V.
  M{\"u}ller.
\newblock Enabling parallel computing in crash.
\newblock {\em Monthly Notices of the Royal Astronomical Society},
  414(1):428--444, 2011.

\bibitem{pease1980reaching}
M. Pease, R. Shostak, and L. Lamport.
\newblock Reaching agreement in the presence of faults.
\newblock {\em Journal of the ACM (JACM)}, 27(2):228--234, 1980.

\bibitem{preparata1981cube}
F.~P. Preparata and J. Vuillemin.
\newblock The cube-connected cycles: a versatile network for parallel
  computation.
\newblock {\em Communications of the ACM}, 24(5):300--309, 1981.

\bibitem{shi:15}
G. Shi, B. Anderson, and U. Helmke.
\newblock Network flows that solve linear equations.
\newblock {\em IEEE Transactions on Automatic Control}, in press, 2017.

\bibitem{tsitsiklis1984distributed}
J.~N. Tsitsiklis, D.~P. Bertsekas, and M. Athans.
\newblock Distributed asynchronous deterministic and stochastic gradient
  optimization algorithms.
\newblock In {\em 1984 American Control Conference}, pages 484--489, 1984.

\bibitem{wang:elia:10}
J.~Wang and N.~Elia.
\newblock Control approach to distributed optimization.
\newblock In {\em Communication, Control, and Computing (Allerton), 2010 48th
  Annual Allerton Conference on}, pages 557--561, Sept 2010.

\bibitem{wang:elia:11}
J.~Wang and N.~Elia.
\newblock A control perspective for centralized and distributed convex
  optimization.
\newblock In {\em 2011 50th IEEE Conference on Decision and Control and
  European Control Conference}, pages 3800--3805, Dec 2011.

\bibitem{yang2014acceleration}
X. Yang and R. Mittal.
\newblock Acceleration of the jacobi iterative method by factors exceeding 100
  using scheduled relaxation.
\newblock {\em Journal of Computational Physics}, 274:695--708, 2014.

\bibitem{young1954iterative}
D. Young.
\newblock Iterative methods for solving partial difference equations of
  elliptic type.
\newblock {\em Transactions of the American Mathematical Society},
  76(1):92--111, 1954.

\end{thebibliography}
\end{document}